\documentclass[aps,prd,twocolumn,showpacs]{revtex4-1}

\usepackage{wallpaper}
\usepackage{amsmath}
\usepackage[colorlinks,plainpages=false]{hyperref}

\begin{document}
%\begin{CJK*}{GBK}{song}

\title{Constraining quark-hadron interface tension in the multi-messenger era}
%\author{Cheng-Jun~Xia (ÏÄîñ¾ý)$^{1}$}
\author{Cheng-Jun~Xia$^{1}$}
\email{cjxia@itp.ac.cn}
\author{Toshiki Maruyama$^{2}$}
\email{maruyama.toshiki@jaea.go.jp}
\author{Nobutoshi Yasutake$^{2, 3}$}
\email{nobutoshi.yasutake@it-chiba.ac.jp}
\author{Toshitaka Tatsumi$^{4}$}
\email{tatsumitoshitaka@gmail.com}

\affiliation{$^{1}${School of Information Science and Engineering, Ningbo Institute of Technology, Zhejiang University, Ningbo 315100, China}
\\$^{2}${Advanced Science Research Center, Japan Atomic Energy Agency, Shirakata 2-4, Tokai, Ibaraki 319-1195, Japan}
\\$^{3}${Department of Physics, Chiba Institute of Technology (CIT), 2-1-1 Shibazono, Narashino, Chiba, 275-0023, Japan}
\\$^{4}${Institute of Education, Osaka Sangyo University, 3-1-1 Nakagaito, Daito, Osaka 574-8530, Japan}}

\date{\today}

\begin{abstract}
We study the interface effects of quark-hadron mixed phase in compact stars. The properties of nuclear matter are obtained
based on the relativistic-mean-field model. For the quark phase, we adopt perturbation model with running quark masses and
coupling constant. At certain choices of parameter sets, it is found that varying the quark-hadron interface tension will
have sizable effects on the radii ($\Delta R \approx 600$ m) and tidal deformabilities ($\Delta \Lambda/\Lambda \approx
50\%$) of {1.36 solar mass} hybrid stars. These provide possibilities for us to constrain the quark-hadron
interface tension with future gravitational wave observations as well as the ongoing NICER mission.
\end{abstract}

%21.65.Qr Quark matter
%25.75.Nq Quark deconfinement, quark-gluon plasma production, and phase transitions
%26.60.Kp Equations of state of neutron-star matter

\maketitle
%\end{CJK*}

\section{\label{sec:intro}Introduction}
With the first observation of gravitational waves from the binary neutron star merger event GW170817~\cite{LVC2017_PRL119-161101,
LVC2019_PRX9-011001}, astrophysics has entered the multi-messenger era. Combined with the electromagnetic observations of the
transient counterpart AT2017gfo and short gamma ray burst GRB170817A~\cite{Coughlin2018_arXiv1812.04803}, the dimensionless combined
tidal deformability of the corresponding compact stars is constrained within $279\leq \tilde{\Lambda}\leq 720$ at 90\% confidence
level~\cite{LVC2017_PRL119-161101, LVC2019_PRX9-011001, Coughlin2018_arXiv1812.04803, Carney2018_PRD98-063004, De2018_PRL121-091102,
Chatziioannou2018_PRD97-104036}. Meanwhile, the recent measurements of neutron stars'
radii indicate that their values lie at the lower end of 10-14 km range~\cite{Guillot2013_ApJ772-7, Lattimer2014_EPJA50-40,
Ozel2016_ARAA54-401, Li2015_ApJ798-56, Steiner2018_MNRAS476-421, Most2018_PRL120-261103, LVC2018_PRL121-161101,
Fattoyev2018_PRL120-172702, Raithel2018_ApJ857-L23}. In light of the precise mass measurements of the two-solar-mass pulsars
PSR J1614-2230 ($1.928 \pm 0.017\ M_\odot$)~\cite{Demorest2010_Nature467-1081, Fonseca2016_ApJ832-167} and PSR J0348+0432
($2.01 \pm 0.04\ M_\odot$)~\cite{Antoniadis2013_Science340-1233232}, we have by far the most stringent constraint
on the equation of states (EoS) of dense matter, which have been examined extensively in the past year~\cite{Zhu2018_ApJ862-98,
Malik2018_PRC98-035804, Sun2019_PRD99-023004, Dexheimer2019_JPG46-034002, Piekarewicz2018_arXiv1812.09974, Gomes2018_arXiv1806.04763,
Sieniawska2019_AA622-A174, Paschalidis2018_PRD97-084038, Han2018_arXiv1810.10967, Montana2018_arXiv1811.10929,
Alvarez-Castillo2019_PRD99-063010, Christian2019_PRD99-023009}.

The situation is even more exciting in the coming years. As the implementation of the {upgraded detectors}, the sensitivity of
gravitational wave observation may be improved by several times, which enables us to observe postmerger signals and constrain
neutron stars' radii to higher accuracy (on the order of a few hundred meters)~\cite{Torres-Rivas2019_PRD99-044014,
Bauswein2019_arXiv1901.06969}. As the X-ray pulse profiles currently being measured by the NICER mission to an unprecedented
accuracy~\cite{Gendreau2016_PSPIE9905-16}, a precise measurement on neutron stars' masses and radii is likely to take place in the
near future~\cite{Zavlinand1998_AA329-583, Bogdanov2007_ApJ670-668, Bogdanov2008_ApJ689-407, Bogdanov2013_ApJ762-96, Ozel2016_ApJ832-92}.
Meanwhile, pulsars that are more massive than PSR J0348+0432 may be expected, e.g.,  {PSR J0740+6620 ($2.17{}_{-0.10}^{+0.11}\
M_\odot$)~\cite{Cromartie2019} and} PSR J2215+5135 ($2.27{}_{-0.15}^{+0.17}\ M_\odot$)~\cite{Linares2018_ApJ859-54}.
Thus, the perspective for future pulsar observations provide opportunities to constrain the properties of dense matter to an
unprecedented accuracy.

At large energy densities, hadronic matter (HM) is expected to undergo a deconfinement phase transition.
For vanishing chemical potentials, a crossover was observed at the critical temperature $T_\mathrm{c}
\approx 170$ MeV~\cite{Aoki2006_Nature443-675}. Similar cases were also expected to occur in dense matter, where the transition
between HM and quark matter (QM) is a smooth crossover~\cite{Baym1979_PA96-131, Celik1980_PLB97-128, Schaefer1999_PRL82-3956,
Fukushim2004_PLB591-277, Hatsuda2006_PRL97-122001, Maeda2009_PRL103-085301, Masuda2013_ApJ764-12, Zhao2015_PRD92-054012,
Kojo2015_PRD91-045003, Masuda2016_EPJA52-65, Whittenbury2016_PRC93-035807}. More traditionally, one expects a first-order
phase transition from HM to QM~\cite{Dexheimer2010_PRC81-045201}{, which provides an important energy source for
the supernova explosion of massive blue supergiant stars~\cite{Fischer2018_NA2-980}.} In such cases, a distinct interface between
quark and hadronic matter
{is} formed. Adopting the Maxwell construction, the properties of hybrid stars with a strong first-order phase
transition and their relevance to gravitational wave observations were investigated~\cite{Sieniawska2019_AA622-A174,
Han2018_arXiv1810.10967, Paschalidis2018_PRD97-084038}. The existence of third family solutions for hybrid stars {was}
examined as well~\cite{Alvarez-Castillo2019_PRD99-063010, Christian2019_PRD99-023009}. It was found that a sharp phase
transition will lead to small tidal deformabilities and induce discontinuities in the relation between tidal deformability
and gravitational mass~\cite{Han2018_arXiv1810.10967}. Meanwhile, a significant deviation from the empirical relation
between the dominant postmerger gravitational wave frequency $f_\mathrm{peak}$ and the radius/tidal deformability of a star
at a given mass was observed if a strong first-order phase transition occurs~\cite{Bauswein2019_PRL122-061102,
Bauswein2019_arXiv1901.06969}. All those
features can {serve} as distinct signals for a strong first-order phase transition in the forthcoming gravitational wave
observations.

Nevertheless, the Maxwell construction for the quark-hadron mixed phase (MP) is only valid if the surface tension $\sigma$
exceeds the critical value $\sigma_\mathrm{c}$~\cite{Maslov2018_arXiv1812.11889}. In fact, depending on the values of $\sigma$,
MP exhibits various structures~\cite{Maruyama2007_PRD76-123015}. The MP consists of point-like HM and QM when the surface
tension $\sigma$ is zero, which is consistent with the Gibbs construction~\cite{Glendenning2000}. If the surface tension value
is moderate, the finite-size effects become important and the geometrical structures such as droplet, rod, slab, tube, and bubble
start to appear~\cite{Heiselberg1993_PRL70-1355, Voskresensky2002_PLB541-93, Tatsumi2003_NPA718-359, Voskresensky2003_NPA723-291,
Endo2005_NPA749-333, Maruyama2007_PRD76-123015, Yasutake2012_PRD86-101302}. The sizes of the geometrical structures increase
with the surface tension and will approach to the limit of Maxwell construction scenarios at $\sigma>\sigma_\mathrm{c}$, i.e.,
bulk separation of quark and hadron phases, which suggests the nonexistence of MP inside hybrid stars.

Such kind of structural differences due to the quark-hadron interface effects are expected to affect many physical processes
in hybrid stars. For example, the coherent scattering of neutrinos off the QM droplets may greatly enhance the neutrino opacity
of the core~\cite{Reddy2000_PLB475-1}. Due to the relaxation of charge neutrality condition, the emergence of hyperons may be
hindered~\cite{Maruyama2008_PLB659-192}, which prevents a fast cooling via the hyperon Urca processes~\cite{Prakash1992_ApJ390-L77,
Tatsumi2003_PTP110-179, Takatsuka2006_PTP115-355}. Despite that the maximum mass of hybrid stars varies little with respect to
the structural differences, it was found that their radii are more affected~\cite{Maruyama2007_PRD76-123015}. Similar cases
were found in Ref.~\cite{Ayriyan2018_PRC97-045802}, where the robustness of third family solutions for hybrid stars was examined
against the formation of pasta structures in the MP. Adopting both the Gibbs and Maxwell constructions for MP, it was shown
that hybrid stars described with the Gibbs construction are more compact and less deformed by the tidal
force~\cite{Gomes2018_arXiv1806.04763, Montana2018_arXiv1811.10929}. Since the Gibbs and Maxwell constructions correspond
to MP obtained at two extreme surface tension values, i.e., $\sigma\rightarrow 0$ and $\sigma>\sigma_\mathrm{c}$, the observations
of neutron stars' radii and tidal deformabilities may provide a unique opportunity to constrain the interface tension $\sigma$.

In the last decade, extensive efforts were made trying to constrain the value of $\sigma$. Based on lattice QCD, the interface
tension was evaluated for vanishing chemical potentials~\cite{Huang1990_PRD42-2864, Huang1991_PRD43-2056, Alves1992_PRD46-3678,
Brower1992_PRD46-2703, Forcrand2005_NPB140-647, Forcrand2005_PRD72-114501}. For dense matter, one has to rely on effective models,
e.g., MIT bag model with color superconductivity~\cite{Oertel2008_PRD77-074015}, linear sigma model~\cite{Palhares2010_PRD82-125018,
Pinto2012_PRC86-025203, Kroff2015_PRD91-025017}, Nambu-Jona-Lasinio model~\cite{Garcia2013_PRC88-025207, Ke2014_PRD89-074041},
three-flavor Polyakov-quark-meson model~\cite{Mintz2013_PRD87-036004}, Dyson-Schwinger equation approach~\cite{Gao2016_PRD94-094030},
{nucleon-meson model~\cite{Fraga2019_PRD99-014046},}
and equivparticle model~\cite{Xia2018_PRD98-034031}, which predict small surface tensions with $\sigma \lesssim 30\
\mathrm{MeV/fm}^{2}$. The quasiparticle model gives slightly larger values for the quark-vacuum interface, i.e., $\sigma= 30 \sim
70\ \mathrm{MeV/fm}^{2}$~\cite{Wen2010_PRC82-025809}. Adopting MRE method, larger surface tension ($\sigma= 145 \sim 165\
\mathrm{MeV/fm}^{2}$) were obtained based on Nambu-Jona-Lasinio model~\cite{Lugones2013_PRC88-045803}, which may vary with
directions in the presence of a strong magnetic field~\cite{Lugones2017_PRC95-015804, Lugones2019_PRC99-035804}. A dimensional
analysis suggests that the surface tension value for color-flavor locked phase may be much larger, e.g., $\sigma \approx 300\
\mathrm{MeV/fm}^{2}$~\cite{Alford2001_PRD64-074017}.

Due to the ambiguities in estimating the values of $\sigma$, in this work we consider the possibilities of constraining $\sigma$
with pulsar observations in the multi-messenger era. In particular, we study the interface effects of quark-hadron mixed phase
in hybrid stars. It is found that the maximum mass, tidal deformabilities, and radii of hybrid stars increase with $\sigma$.
These provide possibilities for us to constrain the quark-hadron interface tension with future gravitational wave observations
as well as the ongoing NICER mission. The paper is organized as follows. In Sec.~\ref{sec:the}, we present our theoretical
framework, where the properties of nuclear matter and quark matter were obtained. The properties of their mixed phases and
the interface effects are investigated in Sec.~\ref{sec:the_MP}, where both the Gibbs and Maxwell constructions are adopted
and examined for the properties of hybrid stars in Sec.~\ref{sec:Gibbs_Maxwell}. As an example, adopting certain choices
of parameters, the geometrical structures in hybrid stars are investigated in Sec.~\ref{sec:pasta}, which verifies our
findings in Sec.~\ref{sec:Gibbs_Maxwell}. Our conclusion is given in Sec.~\ref{sec:con}.

\section{\label{sec:the}Theoretical framework}
\subsection{\label{sec:the_NM} Nuclear matter}
In the mean field approximation, for infinite nuclear matter, the Lagrangian density of relativistic-mean-field
model~\cite{Meng2016_RDFNS} is given as
\begin{eqnarray}
\mathcal{L}
 &=& \sum_{i=n,p} \bar{\psi}_i
       \left\{  i \gamma^\mu \partial_\mu -\gamma^0 \left[g_\omega \omega
               + g_\rho \tau_3 \rho\right] - g_\sigma\sigma \right.
\nonumber \\
 &&\mbox{}\left.  - m_i \right\} \psi_i
     - \frac{1}{2}m_\sigma^2 \sigma^2
     + \frac{1}{2}m_\omega^2 \omega^2
     + \frac{1}{2}m_\rho^2 \rho^2
\nonumber \\
 &&\mbox{}
     +\sum_{i=e,\mu} \bar{\psi}_i \left[ i \gamma^\mu \partial_\mu - m_i\right]\psi_i.  \label{eq:Lagrange}
\end{eqnarray}
Three types of mesons are included to describe the interactions between nucleons, i.e., the isoscalar-scalar meson $\sigma$,
isoscalar-vector meson $\omega$, and isovector-vector meson $\rho$. Note that the coupling constants $g_{\sigma}$, $g_{\omega}$,
and $g_{\rho}$ are density dependent, which were obtained in accordance with the self-energies of Dirac-Brueckner calculations
of nuclear matter~\cite{Typel1999_NPA656-331}, i.e.,
\begin{eqnarray}
g_{\sigma, \omega}(n) &=& g_{\sigma, \omega}(n_0) a_{\sigma, \omega} \frac{1+b_{\sigma, \omega}(n/n_0+d_{\sigma, \omega})^2}
                          {1+c_{\sigma, \omega}(n/n_0+d_{\sigma, \omega})^2}, \label{eq:ddcp_TW}\\
g_{\rho}(n) &=& g_{\rho}(n_0) \exp{\left[-a_\rho(n/n_0-1)\right]}. \label{eq:ddcp_rho}
\end{eqnarray}
Here $n$ is the baryon number density and $n_0$ the saturation density of nuclear matter.

\begin{figure}
\includegraphics[width=\linewidth]{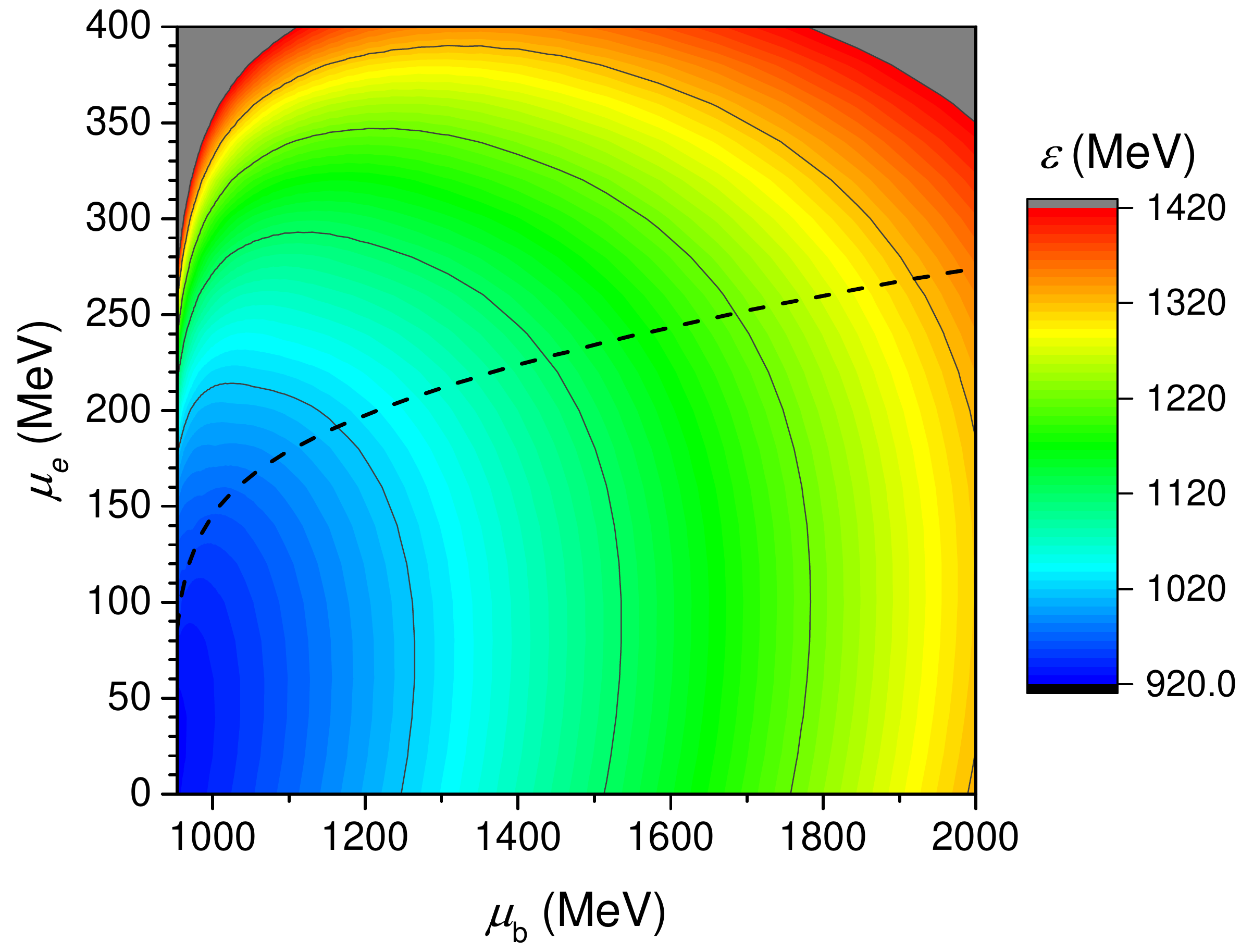}
\caption{\label{Fig:EpA_TW99} Energy per baryon of neutron star matter in $\beta$-equilibrium, which is obtained based on
the effective $N$-$N$ interaction TW99~\cite{Typel1999_NPA656-331}. The dashed curve corresponds to the case where local charge
neutrality condition is satisfied.}
\end{figure}

For the effective $N$-$N$ interactions, we adopt the covariant density functional TW99~\cite{Typel1999_NPA656-331}, which
is consistent with all seven constraints related to symmetric nuclear matter, pure neutron matter, symmetry energy, and
its derivatives~\cite{Dutra2014_PRC90-055203}. Carrying out a standard variational procedure, one obtains the energy
density $E^H$, chemical potential $\mu_i$, and pressure $P^H$ at given particle number densities $n_i$. The energy density
is determined by
\begin{eqnarray}
E^H &=& \sum_{i}\epsilon_i(\nu_i, m_i^*) + \sum_{\phi=\sigma, \omega, \rho} \frac{1}{2}m_\phi^2 \phi^2,
\label{eq:NM_E}
\end{eqnarray}
where $\epsilon_i$ is the kinetic energy density of free Fermi gas at given Fermi momentum $\nu_i$ and effective mass
$m_i^* = m_i + g_{\sigma} \sigma$. Note that the effective masses remain the same for leptons, i.e., $m_{e,\mu}^*\equiv m_{e,\mu}$.
The number density for particle type $i$ is given by $n_i = {\nu_i^3}/{3\pi^2}$, while the chemical potentials for
baryons $\mu_i$ and leptons $\mu_{e,\mu}$ are
\begin{eqnarray}
\mu_i &=& g_{\omega} \omega
              + g_{\rho}\tau_{3} \rho
              + \Sigma^\mathrm{R}
              + \sqrt{\nu_i^2+{m_i^*}^2},
\label{eq:chem_B} \\
\mu_{e,\mu} &=&  \sqrt{\nu_{e,\mu}^2+m_{e,\mu}^2},
\label{eq:chem_l}
\end{eqnarray}
with $\Sigma^\mathrm{R}$ being the ``rearrangement" term, which is introduced to maintain thermodynamic self-consistency
with density dependent coupling constants~\cite{Lenske1995_PLB345-355}. The pressure is obtained with
\begin{equation}
P^H = \sum_i \mu_i n_i  - E^H. \label{eq:NM_P}
\end{equation}

Based on Eqs.~(\ref{eq:NM_E}-\ref{eq:NM_P}), the EoS for nuclear matter is obtained, which gives the saturation density
$n_0=0.153\ \mathrm{fm}^{-3}$, saturation energy $E_0^H/n_0-m_N=-16.25$ MeV, incompressibility $K=240.2$ MeV and symmetry
energy $E_\mathrm{sym}=32.77$ MeV. A detailed contour figure for the energy per baryon $\varepsilon = E^H/n$ of neutron star
matter in $\beta$-equilibrium is presented in Fig.~\ref{Fig:EpA_TW99} as a function of the chemical potentials of baryons
$\mu_\mathrm{b}=\mu_n$ and electrons $\mu_e$, in obtaining which we have disregarded the local charge neutrality condition.

\subsection{\label{sec:the_QM} Quark matter}
At ultra-high densities, the properties of quark matter can be obtained with perturbative QCD (pQCD), which are often extrapolated
to lower density regions~\cite{Fraga2014_ApJ781-L25, Kurkela2014_ApJ789-127}. Similarly, here we adopt pQCD to the order of
$\alpha_\mathrm{s}$ and investigate the properties of quark matter~\cite{Fraga2005_PRD71-105014}, while the non-perturbative
contributions are treated with phenomenological approaches. The pQCD thermodynamic potential density is given by
\begin{equation}
\Omega^\mathrm{pt} = \sum_i^{N_f} \left( \omega^0_i + \omega^1_i \alpha_\mathrm{s} \right), \label{eq:omega}
\end{equation}
with
\begin{eqnarray}
\omega^0_i &=& - \frac{m_i^4}{4 \pi^2}
                 \left[u_i v_i \left( u_i^2 - \frac{5}{2}  \right)
                   + \frac{3}{2} \ln(u_i+v_i)
                 \right],
\label{eq:omega0}\\
\omega^1_i &=& \frac{m_i^4}{2\pi^3}
               \left\{ \left[ 6 \ln\left(\frac{\bar{\Lambda}}{m_i}\right) + 4 \right]\left[u_i v_i - \ln(u_i+v_i)\right] \right.
\nonumber\\
          &&\left.
               + \mbox{} 3\left[u_i v_i - \ln(u_i+v_i)\right]^2 - 2 v_i^4 \right\},
\label{eq:omega1}
\end{eqnarray}
where $u_i \equiv \mu_i/m_i$ and $v_i \equiv \sqrt{u_i^2-1}$ with $\mu_i$ and $m_i$ being
the chemical potential and mass for particle type $i$, respectively. By solving the $\beta$-function and
$\gamma$-function~\cite{Vermaseren1997_PLB405-327} and neglecting higher order terms, the running coupling constant and quark
masses read~\cite{Fraga2005_PRD71-105014}
\begin{eqnarray}
\alpha_\mathrm{s}(\bar{\Lambda})
  &=& \frac{1}{\beta_0 L}   \left(1- \frac{\beta_1\ln{L}}{\beta_0^2 L}\right),
\label{eq:alpha} \\
m_i(\bar{\Lambda})
  &=& \hat{m}_i \alpha_\mathrm{s}^{\frac{\gamma_0}{\beta_0}}
      \left[ 1 + \left(\frac{\gamma_1}{\beta_0}-\frac{\beta_1\gamma_0}{\beta_0^2}\right) \alpha_\mathrm{s} \right].
\label{eq:mi}
\end{eqnarray}
Here $L=2 \ln\left( \frac{\bar{\Lambda}}{\Lambda_{\overline{\mathrm{MS}}}}\right)$ with $\Lambda_{\overline{\mathrm{MS}}} = 376.9$ MeV
being the $\overline{\mathrm{MS}}$ renormalization point, while the invariant quark masses are $\hat{m}_u= 3.8$ MeV, $\hat{m}_d = 8$ MeV,
and $\hat{m}_s = 158$ MeV according to the results obtained by Particle Data Group~\cite{Olive2014_CPC38-090001}. Note that
$\beta_0=\frac{1}{4\pi}(11-\frac{2}{3}N_\mathrm{f})$ and $\beta_1=\frac{1}{16 \pi^2} (102-\frac {38}{3} N_\mathrm{f})$ for the
$\beta$-function while $\gamma_0=1/\pi$ and $\gamma_1=\frac{1}{16\pi^2} (\frac{202}{3} - \frac{20}{9}N_\mathrm{f})$ for the $\gamma$-function.
At present, it is not clear how the renormalization scale $\bar{\Lambda}$  evolves with the chemical potentials of quarks, where many
possibilities exist~\cite{Fraga2005_PRD71-105014, Xu2015_PRD92-025025}. In this work, we adopt the following formalism:
\begin{equation}
  \bar{\Lambda} = \frac{C}{3} \sum_{i=u,d,s}\mu_i, \label{eq:Lambda}
\end{equation}
with $C=1 \sim 4$~\cite{Fraga2014_ApJ781-L25}.

To incorporate the non-perturbative effects, we introduce an extra bag constant $B$ to take into account the energy difference between the
physical and perturbative vacua. According to various studies, it was found that the bag constant can vary with state variables, e.g.,
the temperature~\cite{Song1992_PRD46-3211, Gorenstein1995_PRD52-5206}, chemical potentials of quarks~\cite{Gardim2009_NPA825-222},
density~\cite{Burgio2002_PLB526-19} and even magnetic field~\cite{Wen2012_PRD86-034006}. The bag constant at vanishing chemical potentials
is found to be around $455\ \mathrm{MeV\ fm}^{-3}$ according to QCD sum-rule~\cite{Shuryak1978_PLB79-135}, while carrying out fits to light
hadron spectra suggests $B\approx50\ \mathrm{MeV\ fm}^{-3}$~\cite{DeGrand1975_PRD12-2060}. At larger chemical potentials, comparing
Eq.~(\ref{eq:omega}) with the pQCD calculations to the order of $\alpha_\mathrm{s}^2$~\cite{Fraga2014_ApJ781-L25}, an increasing difference
on the thermodynamic potential density is observed. At the same time, the dynamic equilibrium condition at the critical temperature
of deconfinement phase transition demands $B \approx 400\ \mathrm{MeV\ fm}^{-3}$~\footnote{Obtained by equating the pressures of QGP
($-B+37\pi^2 T^4/90$) and pion gas ($\pi^2 T^4/30$) at $T = T_c$ ($\sim$170 MeV).}, indicating a large bag constant value at high
energy density. On combination of those values, similar to Ref.~\cite{Burgio2002_PLB526-19, Maieron2004_PRD70-043010}, we adopt the
following parametrization of $B$, i.e.,
\begin{equation}
B = B_\mathrm{QCD} + (B_0 - B_\mathrm{QCD})
    \exp{\left[-\left( \frac{\sum_i\mu_i-930}{\Delta\mu}\right)^4\right]},
\label{eq:BL}
\end{equation}
which gives $B = B_0 = 50\ \mathrm{MeV\ fm}^{-3}$ at $\mu_u+\mu_d+\mu_s = 930$ MeV. The width parameter $\Delta\mu$ and
$B_\mathrm{QCD}$ are left undetermined and to be fixed later. Note that adopting smaller $B_\mathrm{QCD}$ reduces the maximum
mass of hybrid stars and shrinks the parameter space for $\Delta\mu$ and $C$ in light of the observational mass of PSR
J0348+0432~\cite{Antoniadis2013_Science340-1233232}, which is indicated in Fig.~\ref{Fig:Mmax_B}.

Combining both the pQCD results in Eq.~(\ref{eq:omega}) and parameterized bag constant in Eq.~(\ref{eq:BL}), the thermodynamic potential
density for quark matter is obtained with $\Omega^Q =  \Omega^\mathrm{pt} + \omega^0_e/3 + B$, including the contributions of electrons.
Based on the basic thermodynamic relations, the particle number density is $n_i = - \frac{\partial \Omega}{\partial \mu_i}$,
and energy density of quark matter
\begin{equation}
E^Q =  \Omega^\mathrm{pt} + \frac{1}{3}\omega^0_e +  B + \sum_i \mu_i n_i. \label{eq:Eq}
\end{equation}
The pressure takes negative values of the thermodynamic potential density, i.e., $P^Q=-\Omega^Q$.

\subsection{\label{sec:EoS} Approximate the EoSs of HM and QM}
For matter inside compact stars, to reach the lowest energy, particles will undergo weak reactions until the $\beta$-equilibrium condition
is fulfilled, i.e.,
\begin{equation}
\mu_i= B_i \mu_\mathrm{b} - q_i \mu_e,  \label{eq:weakequi}
\end{equation}
where $B_i$ ($B_p=B_n=1$, $B_u=B_d=B_s=1/3$, and $B_e=B_\mu=0$) is the baryon number and $q_i$ ($q_p=1$, $q_n=0$, $q_u=2/3$, $q_d=q_s=-1/3$
and $q_e=q_\mu=-1$) the charge of particle type $i$. Note that the chemical potential of neutrinos is set to zero since they can leave the
system freely.

\begin{figure}
\includegraphics[width=\linewidth]{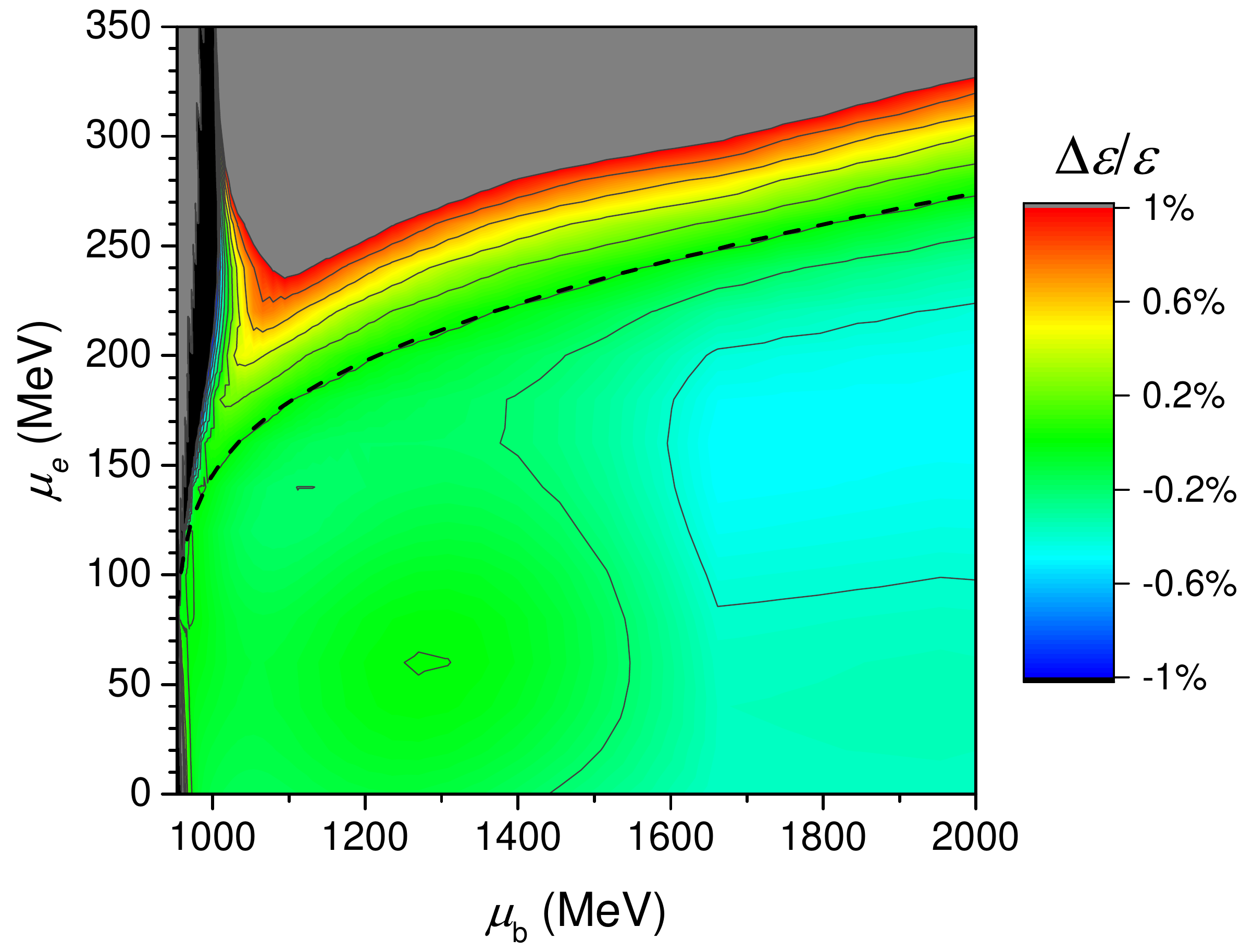}
\caption{\label{Fig:DEpA_TW99} The relative deviations of the obtained energy per baryon from those of Fig.~\ref{Fig:EpA_TW99}.}
\end{figure}

To simplify our calculation, it is convenient to approximate the pressures and energy densities of HM and QM by expanding
them with respect to $\mu_e$, i.e.,
\begin{eqnarray}
P(\mu_\mathrm{b}, \mu_e) &=& P_0(\mu_\mathrm{b}) - \frac{1}{2} n_\mathrm{ch}'(\mu_\mathrm{b}) [\mu_e - \mu_{e0}(\mu_\mathrm{b})]^2,
\label{eq:P_lin} \\
E(\mu_\mathrm{b}, \mu_e) &=& E_0(\mu_\mathrm{b}) + E'(\mu_\mathrm{b}) [\mu_e - \mu_{e0}(\mu_\mathrm{b})] \nonumber \\
&&{}+ \frac{1}{2} E''(\mu_\mathrm{b}) [\mu_e - \mu_{e0}(\mu_\mathrm{b})]^2. \label{eq:E_lin}
\end{eqnarray}
Here $P_0$, $E_0$, and $\mu_{e0}$ is the pressure, energy density, and electron chemical potential obtained by fulfilling the local
charge neutrality condition $n_\mathrm{ch} = \sum q_i n_i = 0$. We have adopted prime notion to represent taking derivatives with
respect to $\mu_e$ at $\mu_e=\mu_{e0}$, i.e.,
\begin{equation}
n_\mathrm{ch}' =\frac{\partial n_\mathrm{ch}}{\partial \mu_e},~~E' = \frac{\partial E}{\partial \mu_e},~~E'' =
\frac{\partial^2 E}{\partial \mu_e^2}. \nonumber
\end{equation}
Note that $n_\mathrm{ch}'$ is related to the Debye screening length with $\lambda_\mathrm{D}\equiv \left( -4\pi \alpha n_\mathrm{ch}'\right)^{-1/2}$.
Based on Eqs.~(\ref{eq:P_lin}-\ref{eq:E_lin}) and basic thermodynamic relations, we have
\begin{eqnarray}
n_\mathrm{ch}(\mu_\mathrm{b}, \mu_e) &=& -\left.\frac{\partial P}{\partial \mu_e}\right|_{\mu_\mathrm{b}} = n_\mathrm{ch}' (\mu_e - \mu_{e0}),
\label{eq:nch} \\
n(\mu_\mathrm{b}, \mu_e) &=& (E + \mu_e n_\mathrm{ch} + P)/\mu_\mathrm{b}. \label{eq:n}
\end{eqnarray}
The obtained properties of HM and QM in Sec.~\ref{sec:the_NM} and Sec.~\ref{sec:the_QM} are then well reproduced by Eqs.~(\ref{eq:P_lin}-\ref{eq:n}).
As an example,
in Fig.~\ref{Fig:DEpA_TW99} we plot the relative deviations of energy per baryon for nuclear matter with $\Delta\varepsilon =
\varepsilon^\mathrm{cal} - \varepsilon^\mathrm{fit}$, which lies within 1\%.

\section{\label{sec:the_MP} Mixed phase and interface effects}

\subsection{\label{sec:Gibbs_Maxwell} The Gibbs and Maxwell constructions}
To investigate the effects of quark-hadron interface on the properties of MP and compact stars, we consider two extreme cases, i.e.,
the Gibbs construction at $\sigma\rightarrow 0$ and the Maxwell construction at $\sigma>\sigma_\mathrm{c}$. In both cases, at a
given baryon chemical potential $\mu_\mathrm{b}$, the dynamic stability condition needs to be satisfied, i.e.,
\begin{equation}
P^H = P^Q. \label{eq:dyn_stb}
\end{equation}

In principle, leptons are free to move throughout the quark-hadron interface, then the chemical potentials of electrons in each
phase become the same, i.e., $\mu_e^H = \mu_e^Q$, which is the case for the Gibbs construction. For the Maxwell construction, the
scale of MP is much larger than the Debye screening length $\lambda_\mathrm{D}$, so that the local charge neutrality condition
is effectively restored due to Coulomb repulsion. Thus, for the two types of phase construction schemes, we have
\begin{eqnarray}
\mathrm{Gibbs:}&&   ~~\mu_e^H = \mu_e^Q,    ~~(1-\chi) n_\mathrm{ch}^H + \chi n_\mathrm{ch}^Q = 0; \label{eq:glb_ch} \\
\mathrm{Maxwell:}&& ~~\mu_e^H \neq \mu_e^Q, ~~n_\mathrm{ch}^H =0, ~~n_\mathrm{ch}^Q = 0. \label{eq:loc_ch}
\end{eqnarray}
Here the quark fraction $\chi\equiv V^Q/V$ with $V^Q$ being the volume occupied by quarks and $V$ the total volume. Based on
Eqs.~(\ref{eq:P_lin}-\ref{eq:n}), Eqs.~(\ref{eq:dyn_stb}-\ref{eq:loc_ch}) can be solved analytically at given $\mu_\mathrm{b}$.
Then the properties of MP can be obtained.

\begin{figure}
\includegraphics[width=\linewidth]{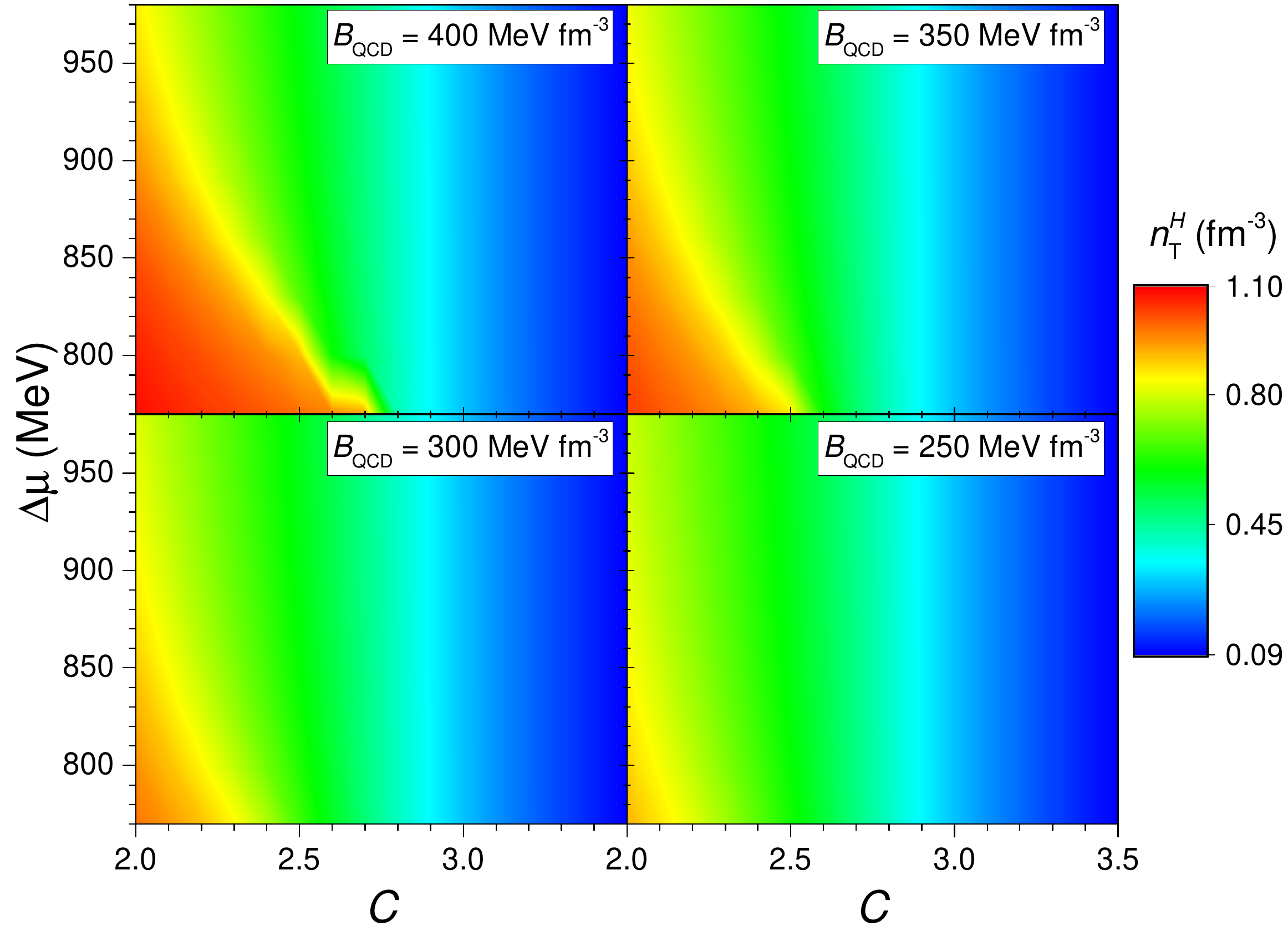}
\caption{\label{Fig:nHT} The density of nuclear matter on the occurrence of deconfinement phase transition {obtained with
the Maxwell construction}, beyond which quark matter start to appear.}
\end{figure}

\begin{figure}
\includegraphics[width=\linewidth]{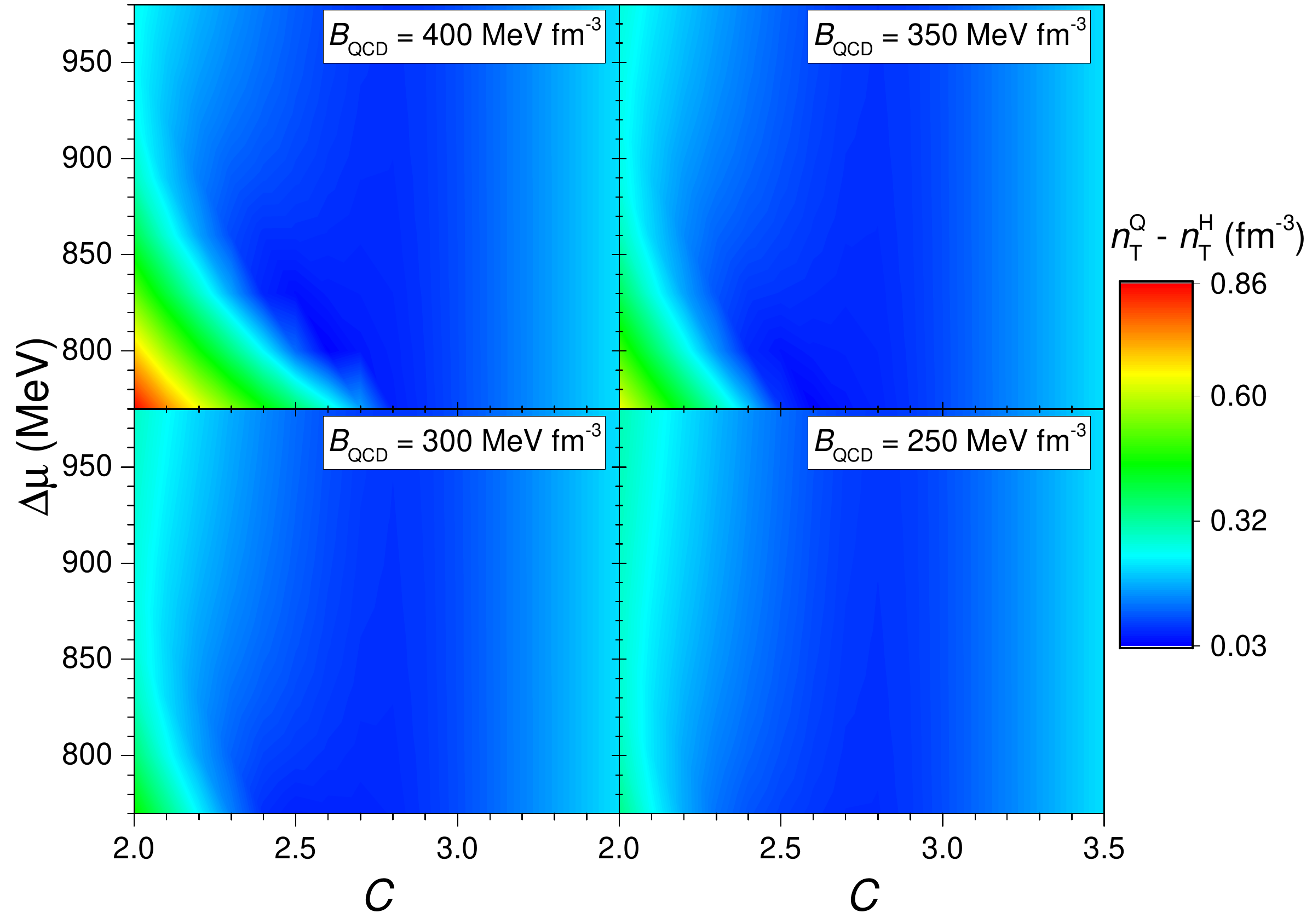}
\caption{\label{Fig:DnT} The difference between the densities of nuclear matter and quark matter on the occurrence of deconfinement
phase transition {which is obtained with the Maxwell construction}.}
\end{figure}

Adopting both the Gibbs and Maxwell constructions, we investigate the properties of MP at various parameter sets with $C=2 \sim 3.5$
and $\Delta\mu=770 \sim 1000$ MeV. Note that the deconfinement phase transition occurs at densities smaller than $0.09\ \mathrm{fm}
^{-3}$ if $C \gtrsim 3.5$, while at $\Delta\mu \lesssim 770$ MeV and $B_\mathrm{QCD} = 400\ \mathrm{MeV\ fm}^{-3}$ the velocity
of sound in QM may exceeds the speed of light, which are excluded in our calculation. By solving Eq.~(\ref{eq:dyn_stb}) and
(\ref{eq:loc_ch}), the densities of nuclear matter $n^H_\mathrm{T}$ and quark matter $n^Q_\mathrm{T}$ on the occurrence of
deconfinement phase transition can be obtained at given $C$, $\Delta\mu$, and $B_\mathrm{QCD}$, which are presented in
Figs.~\ref{Fig:nHT} and \ref{Fig:DnT}. Since the energy per baryon of QM decreases if we adopt larger $C$, $\Delta\mu$, and smaller
$B_\mathrm{QCD}$, the transition density $n^H_\mathrm{T}$ decreases accordingly. The density jump $n^Q_\mathrm{T}-n^H_\mathrm{T}$
is increasing with $B_\mathrm{QCD}$ and decreasing with $C$ and $\Delta\mu$, where a large $n^Q_\mathrm{T}-n^H_\mathrm{T}$
indicates a strong first-order phase transition. At $C\gtrsim 2.8$, we find varying $\Delta\mu$ or $B_\mathrm{QCD}$ does
not affect the transition densities $n^H_\mathrm{T}$ and $n^Q_\mathrm{T}$, while $n^Q_\mathrm{T}$ decreases slightly with $C$.

\begin{figure}
\includegraphics[width=\linewidth]{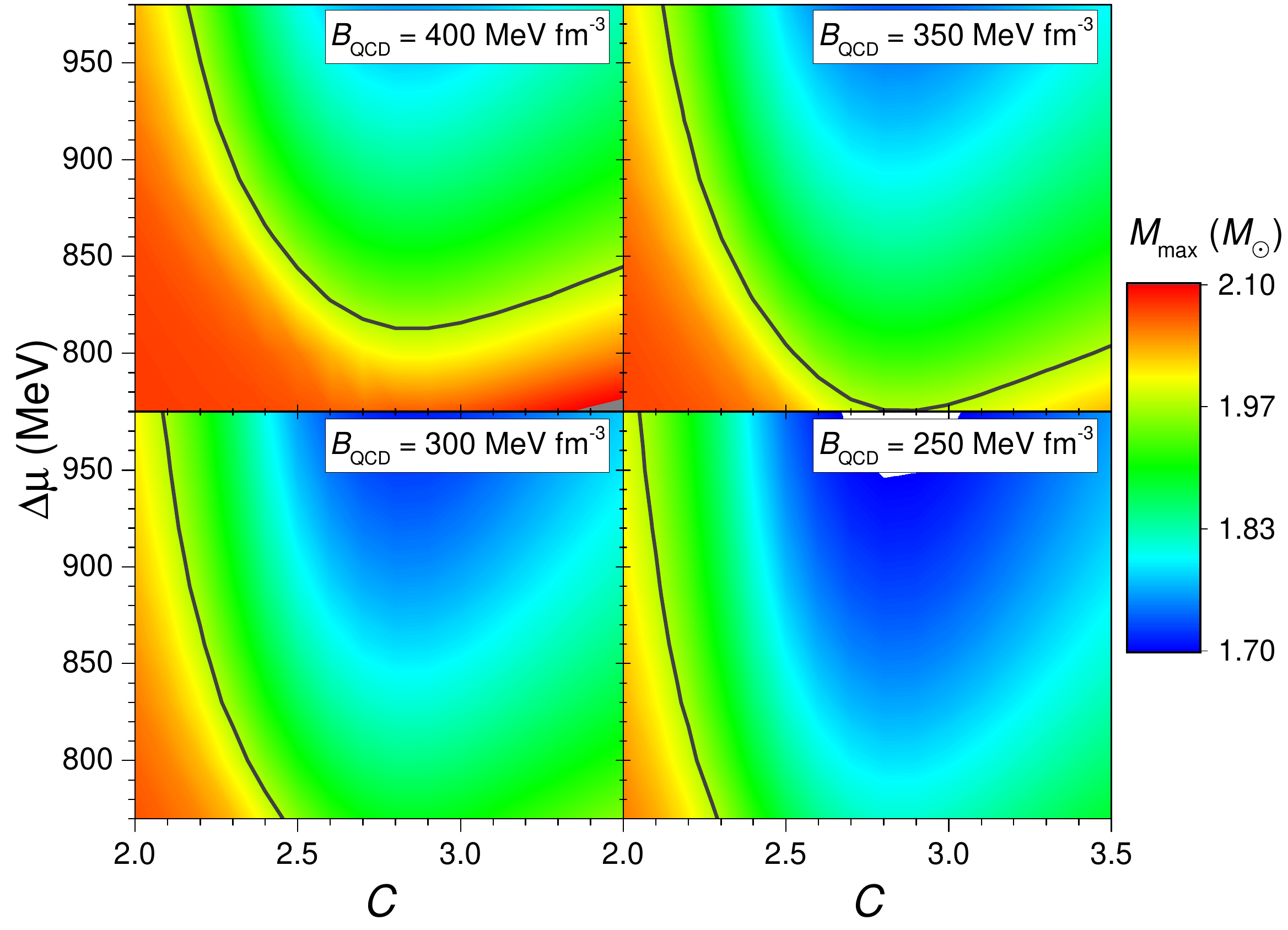}
\caption{\label{Fig:Mmax_B} The maximum mass of hybrid stars obtained with the Maxwell construction. The solid curves
correspond to the cases with $M_\mathrm{max} = 1.97\ M_\odot$, i.e., the lower limit of the observational mass of PSR
J0348+0432~\cite{Antoniadis2013_Science340-1233232}.}
\end{figure}

\begin{figure}
\includegraphics[width=\linewidth]{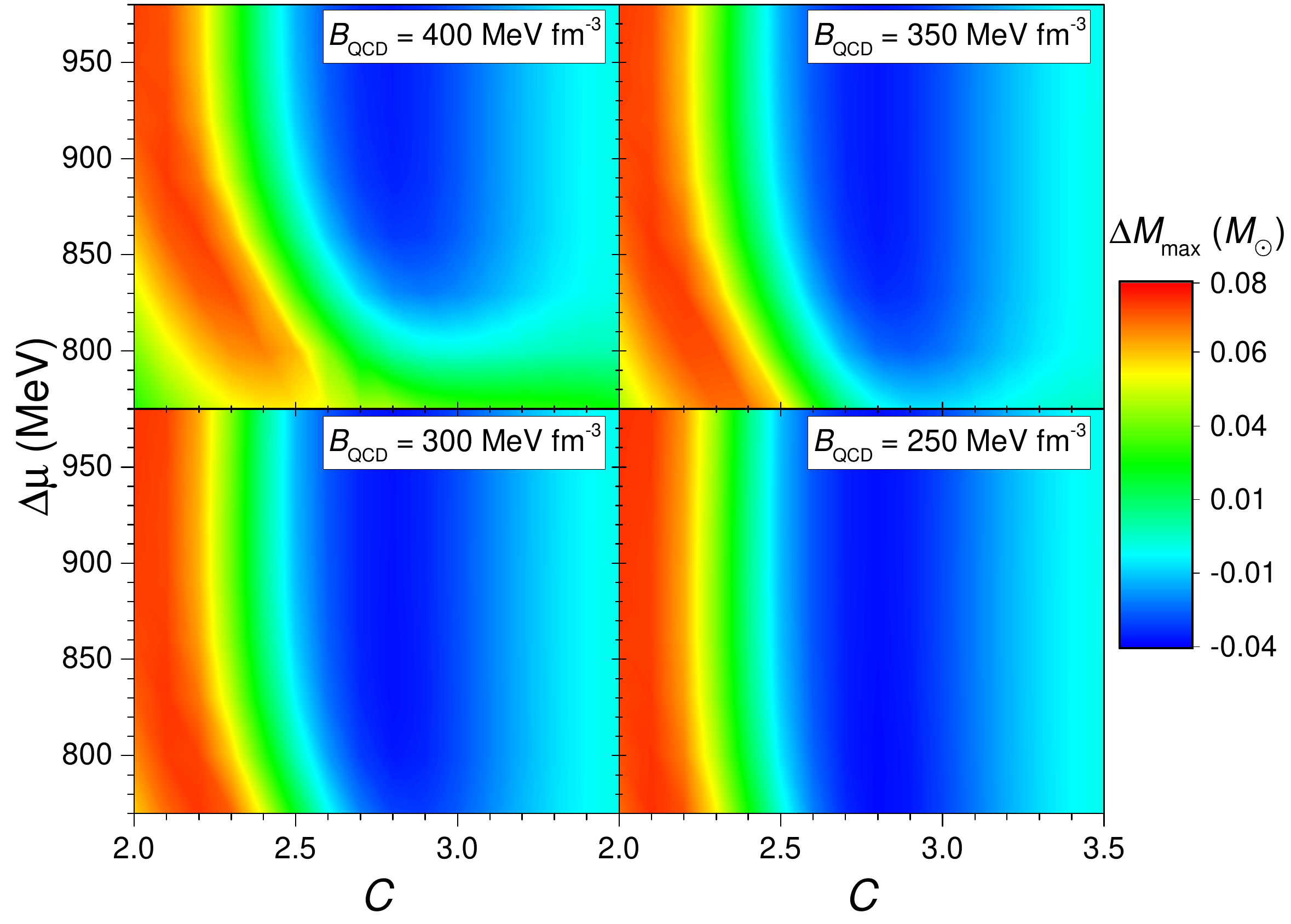}
\caption{\label{Fig:DMmax_B} The variations on the maximum mass of hybrid stars caused by introducing the Gibbs and
Maxwell constructions.}
\end{figure}

Finally, based on the EoSs of NM, QM, and MP, we solve the Tolman-Oppenheimer-Volkov (TOV) equation
\begin{equation}
\frac{\mbox{d}P}{\mbox{d}r}
=-\frac{G M E}{r^2}
  \frac{(1+P/E)(1+4\pi r^3 P/M)} {1-2G M/r}  \label{eq:TOV}
\end{equation}
with subsidiary condition
\begin{equation}
\frac{\mbox{d}M(r)}{\mbox{d}r} = 4\pi E r^2. \label{eq:m_star}
\end{equation}
Here the gravity constant is taken as $G=6.707\times 10^{-45}\ \mathrm{MeV}^{-2}$. Note that at subsaturation densities,
uniform nuclear matter becomes unstable and geometrical structures emerge. In such cases, we adopt the EoS presented
in Refs.~\cite{Feynman1949_PR75-1561, Baym1971_ApJ170-299, Negele1973_NPA207-298} at $n<0.08\ \mathrm{fm}^{-3}$.
The mass $M$ and radius $R$ of a compact star are obtained at given centre pressure. In Fig.~\ref{Fig:Mmax_B} we present
the maximum mass $M_\mathrm{max}$ of hybrid stars obtained with the Maxwell construction. It is found that $M_\mathrm{max}$
decreases with $\Delta\mu$ and increases with $B_\mathrm{QCD}$. At fixed $\Delta\mu$ and $B_\mathrm{QCD}$, the obtained
maximum mass decreases with {increasing} $C$ at $C\lesssim 2.8$. This is mainly due to the softening of EoSs with the occurrence
of deconfinement phase transition. For larger $C$, as indicated in Fig.~\ref{Fig:nHT}, QM appears at densities smaller
than $2 n_0$. In such cases, the core of a hybrid star is comprised almost entirely of QM, which has a similar parameter
dependence on $C$ as a strange star~\cite{Xia2017_NPB916-669}, i.e., the corresponding $M_\mathrm{max}$ is increasing
with $C$. For our calculation to be consistent with the observational mass of PSR J0348+0432 ($2.01 \pm 0.04\
M_\odot$)~\cite{Antoniadis2013_Science340-1233232}, smaller $\Delta\mu$, $C$, and larger $B_\mathrm{QCD}$ are favored,
i.e., the lower left regions in Fig.~\ref{Fig:Mmax_B} with $M_\mathrm{max}>1.97\ M_\odot$. This area in the parameter
space shrinks if we adopt smaller $B_\mathrm{QCD}$. Note that introducing the Gibbs construction will result
in a different maximum mass. In Fig.~\ref{Fig:DMmax_B} we present the variations on the maximum mass of hybrid stars
$\Delta M_\mathrm{max} =M_\mathrm{max}^\mathrm{Maxwell} - M_\mathrm{max}^\mathrm{Gibbs}$ caused by introducing the Gibbs and
Maxwell constructions. It is found that the difference is larger and positive at smaller $C$, while $\Delta M_\mathrm{max}$
becomes negative and approaches to its minimum at $C\approx 2.8$. Nevertheless, $\Delta M_\mathrm{max}$ is positive
for the cases with $M_\mathrm{max}>1.97\ M_\odot$, where the obtained $M_\mathrm{max}^\mathrm{Maxwell}$ is larger than
$M_\mathrm{max}^\mathrm{Gibbs}$. In general, we find that the difference is insignificant
($|\Delta M_\mathrm{max}|\lesssim 0.08\ M_\odot$) comparing with the masses of hybrid stars.

\begin{figure}
\includegraphics[width=\linewidth]{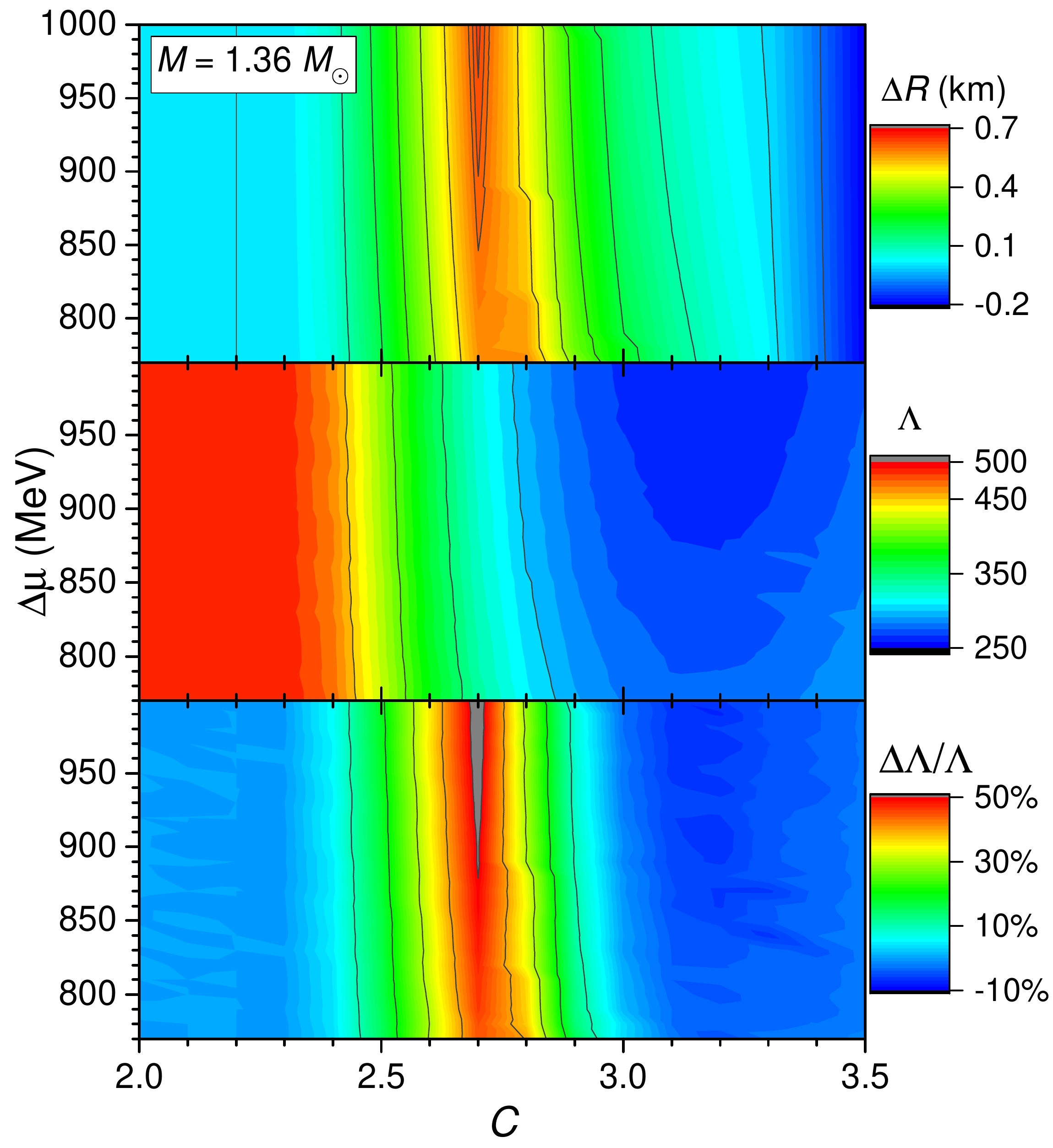}
\caption{\label{Fig:M136RL} The tidal deformability (centre panel) of hybrid stars at $M=1.36\ M_\odot$ obtained with
the Gibbs construction. The variations on the radius (top panel) and tidal deformability (bottom panel) of hybrid stars
caused by introducing the Gibbs and Maxwell constructions. Here we take $B_\mathrm{QCD} = 400\ \mathrm{MeV\ fm}^{-3}$.}
\end{figure}

The tidal deformability can be estimated with
\begin{equation}
\Lambda = \frac{2 k_2}{3}\left( \frac{R}{G M} \right)^5, \label{eq:td}
\end{equation}
where $k_2$ is the second Love number~\cite{Damour2009_PRD80-084035, Hinderer2010_PRD81-123016, Postnikov2010_PRD82-024016}.
For hybrid stars with $M=1.36\ M_\odot$, it is found that both tidal deformability $\Lambda_{1.36}$ and radius $R_{1.36}$
are insensitive to the choices of $B_\mathrm{QCD}$ according to our calculation. We thus take $B_\mathrm{QCD} = 400\
\mathrm{MeV\ fm}^{-3}$ and present the obtained $\Lambda_{1.36}$ with the Gibbs construction at the centre panel of
Fig.~\ref{Fig:M136RL}, which is decreasing with $C$ but insensitive to $\Delta\mu$ and $B_\mathrm{QCD}$. If we assume
the mass ratio $m_2/m_1 = 1$ for the binary neutron star merger event GW170817, combined with the measured chirp mass
$\mathcal{M} = {(m_1 m_2)^{3/5}}{(m_1+m_2)^{-1/5}}=1.186\pm 0.001\ M_\odot$~\cite{LVC2019_PRX9-011001}, we then have
$m_1=m_2=1.362\ M_\odot$ and the dimensionless combined tidal deformability $\tilde{\Lambda}=\Lambda_1=\Lambda_2 \approx
\Lambda_{1.36}$ with the constraint $279\leq \tilde{\Lambda} \leq 720$~\cite{LVC2017_PRL119-161101, LVC2019_PRX9-011001,
Coughlin2018_arXiv1812.04803, Carney2018_PRD98-063004, De2018_PRL121-091102, Chatziioannou2018_PRD97-104036}. In fact,
the obtained $\tilde{\Lambda}$ may deviate slightly from $\Lambda_{1.36}$ for other mass ratios as indicated in
Fig.~\ref{Fig:LambdaT}, while the variations are insignificant. In such cases, the region in the parameter space centered
at $C\approx 3.2$ can be excluded since {a lower limit with $\tilde{\Lambda} > 279$  was obtained based on the
Bayesian analysis of the combined information from GW170817, AT2017gfo, and GRB170817~\cite{Coughlin2018_arXiv1812.04803}}.
To show the interface effects on the properties
of hybrid stars at $M=1.36\ M_\odot$, we compare the radii and tidal deformabilities of hybrid stars obtained based
on the Gibbs and Maxwell constructions. The variations on $\Lambda_{1.36}$ and $R_{1.36}$ are presented in the top and
bottom panels of Fig.~\ref{Fig:M136RL}, where $\Delta R = R^\mathrm{Maxwell} - R^\mathrm{Gibbs}$ and $\Delta
\Lambda/\Lambda = \Lambda^\mathrm{Maxwell}/\Lambda^\mathrm{Gibbs} - 1$. At certain choice of parameters, e.g.,
$C\approx 2.7$, the interface effects on the properties of hybrid stars become sizable. It is found that the radius
of a hybrid star at $M=1.36\ M_\odot$ may vary up to $600$ m, which is within the capability of the NICER
mission~\cite{Gendreau2016_PSPIE9905-16} or gravitational wave observations~\cite{Torres-Rivas2019_PRD99-044014,
Bauswein2019_arXiv1901.06969}. Meanwhile, the relative variations on the tidal deformability may even reach $50\%$,
which can be distinguished by future gravitational wave observations. {Nevertheless, it is worth mentioning
that for traditional neutron stars without a deconfinement phase transition, the properties of nuclear matter at high
densities would have sizable impacts on the radii and tidal deformabilities as well, e.g., the symmetry energy
slope~\cite{Zhu2018_ApJ862-98, Dexheimer2019_JPG46-034002}. In such cases, due to the uncertainties in the properties
of hadronic matter, it is necessary to adopt other hadronic EoSs in our future study and examine their impacts, e.g.,
DD2 EoS with light clusters~\cite{Typel2010_PRC81-015803} or the nuclear EoS predicated by the cluster variational method
using the Jastrow wave function~\cite{Togashi2017_NPA961-78}.}

\begin{figure}
\includegraphics[width=\linewidth]{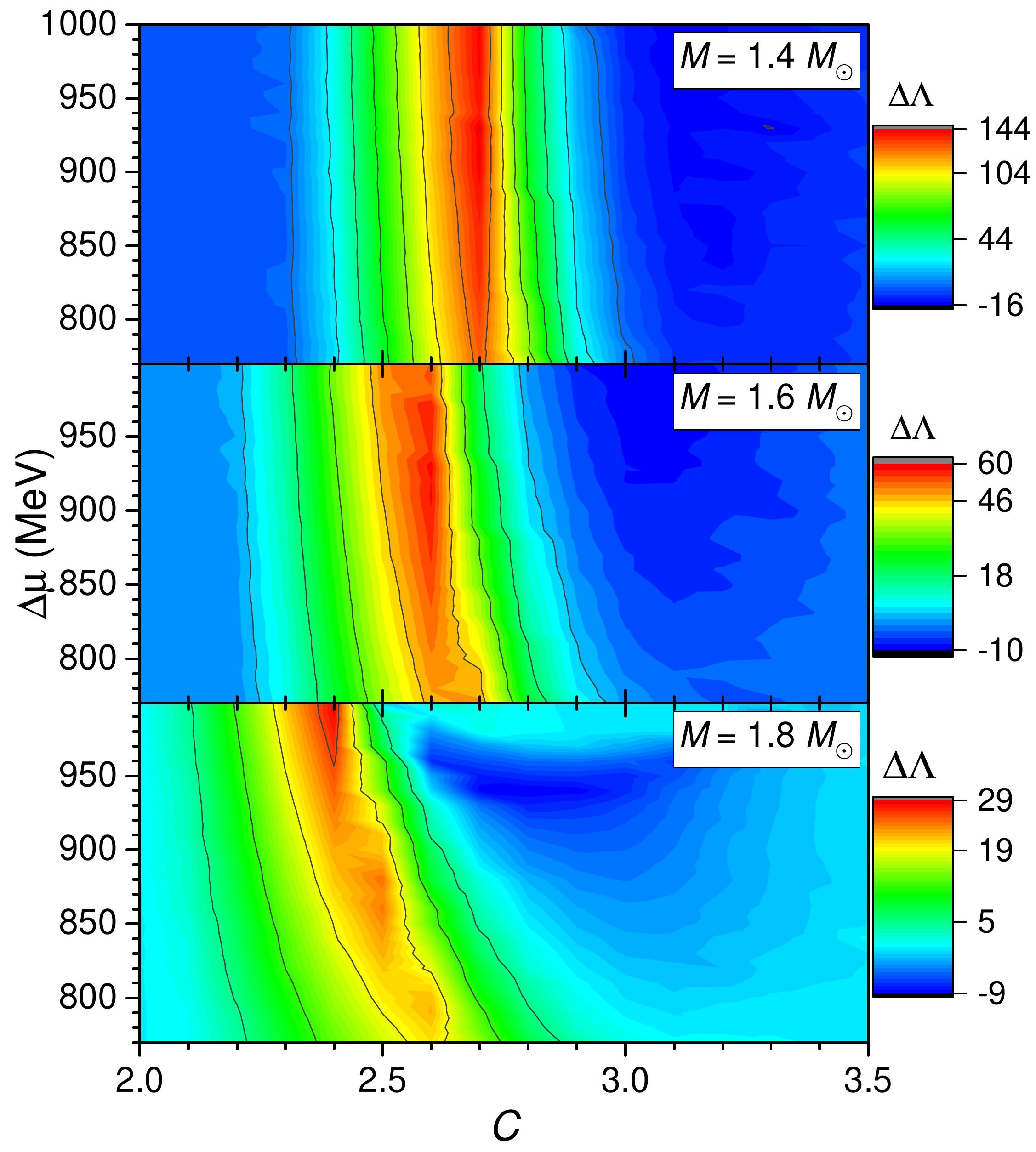}
\caption{\label{Fig:DL1468} The variations on the tidal deformability of hybrid stars at $M=1.4$, 1.6, and $1.8\ M_\odot$
caused by introducing the Gibbs and Maxwell constructions, where $B_\mathrm{QCD} = 400\ \mathrm{MeV\ fm}^{-3}$ is adopted.}
\end{figure}

To further examine the interface effects on more massive hybrid stars, in Fig.~\ref{Fig:DL1468} we present the variations
of tidal deformability with $M=1.4$, 1.6, and $1.8\ M_\odot$ obtained at $B_\mathrm{QCD} = 400\ \mathrm{MeV\ fm}^{-3}$.
It is found that the region with large $\Delta\Lambda$ in the parameter space varies with the mass of hybrid stars,
where the centre shifts to smaller $C$ as $M$ increases. This is mainly due to the fact that the interface effects
become important when deconfinement phase transition starts to take place at the centre of the star, which is around
the densities indicated in Figs.~\ref{Fig:nHT} and \ref{Fig:DnT}. Similar cases are expected for the radii of hybrid
stars as well as adopting other values of $B_\mathrm{QCD}$.

In summary, based on the results indicated in Figs.~\ref{Fig:nHT}-\ref{Fig:DL1468}, the parameter $C$ can be constrained
and is likely small ($\lesssim3$) according to the expected hadron-quark transition density in heavy-ion collision
phenomenology, the observational mass of PSR J0348+0432~\cite{Antoniadis2013_Science340-1233232}, and the lower limit of
the dimensionless combined tidal deformability~\cite{Coughlin2018_arXiv1812.04803}. In such cases, the interface effects play important
roles for the radii and tidal deformabilities of hybrid stars as indicated in Figs.~\ref{Fig:M136RL} and \ref{Fig:DL1468}.
With the {upgraded} gravitational wave detectors~\cite{Torres-Rivas2019_PRD99-044014,
Bauswein2019_arXiv1901.06969}, the ongoing NICER mission~\cite{Gendreau2016_PSPIE9905-16}, and the mass measurements
of massive pulsars~\cite{Linares2018_ApJ859-54}, we may have a good chance to constrain simultaneously the parameters $C$,
$\Delta\mu$, $B_\mathrm{QCD}$, as well as the quark-hadron interface tension in the near future with the accurately
measured masses, radii, and tidal deformabilities of pulsars.

\subsection{\label{sec:pasta} Geometrical structures}

Since the emergence of geometrical structures is inevitable if the interface tension $\sigma<\sigma_\mathrm{c}$, it is
necessary to investigate the interface effects on those structures and consequently on the properties of MP. To construct
the geometrical structures of MP, we employ a Wigner-Seitz approximation and assume spherical symmetry, i.e., only the
droplet and bubble phases are considered.

As was done in Refs.~\cite{Xia2016_SciBull61-172, Xia2016_SciSinPMA46-012021_E, Xia2016_PRD93-085025, Xia2017_NPB916-669}
but neglecting the contributions of gravity, the internal structure of the Wigner-Seitz cell is determined by minimizing
the mass, which is consistent with the constancy of chemical potentials
\begin{equation}
\bar{\mu}_i = \mu_i(r) + q_i \varphi(r) = \mathrm{constant}, \label{eq:pdis}
\end{equation}
with the electric potential $\varphi(r)$ determined by
\begin{equation}
r^2\frac{\mbox{d}^2\varphi}{\mbox{d}r^2} + 2r \frac{\mbox{d}\varphi}{\mbox{d}r} + 4\pi\alpha r^2 n_\mathrm{ch}(r) = 0. \label{eq:efield}
\end{equation}
Here $\alpha=1/137$ is the fine-structure constant.
Since the $\beta$-equilibrium condition is fulfilled, the local chemical potentials are determined by Eq.~(\ref{eq:weakequi}) with a
constant $\mu_\mathrm{b}$ and space dependent $\mu_e(r) = \varphi(r) + m_e$. With the linearization adopted in Eq.~(\ref{eq:nch}),
Eq.~(\ref{eq:efield}) can be solved analytically and gives
\begin{eqnarray}
\varphi^\mathrm{I} &=& \frac{C^\mathrm{I}}{r} \sinh\left( \frac {r}{ \lambda_\mathrm{D}^\mathrm{I} } \right)+\varphi^\mathrm{I}_0,  \label{eq:phi_I} \\
\varphi^\mathrm{O} &=& \frac{C^\mathrm{O}}{r \left( R_\mathrm{W} +\lambda_\mathrm{D}^\mathrm{O} \right) }
                       \left[ \sinh(\tilde{r}) \lambda_\mathrm{D}^\mathrm{O} + \cosh(\tilde{r}) R_\mathrm{W} \right]+\varphi^\mathrm{O}_0 \nonumber \\
&& \mathrm{with} ~~ \tilde{r}\equiv{(r-R_\mathrm{W})}/{\lambda_\mathrm{D}^\mathrm{O}}. \label{eq:phi_O}
\end{eqnarray}
Here the Wigner-Seitz cell is divided into the inner part (I) and outer part (O), i.e., a small sphere with radius $R$ enclosed
within a spherical shell with outer radius $R_\mathrm{W}$. The MP is at the droplet phase if we have QM located at the inner part
and HM in the outer part, and vice versa, the MP is at the bubble phase.
The electric fields $\varphi^\mathrm{I}(r)$ ($r<R$) and $\varphi^\mathrm{O}(r)$ ($R<r\leq R_\mathrm{W}$) and their derivatives should
match with each other at $r=R$, which determines the parameters $C^\mathrm{I}$, $\varphi^\mathrm{I}_0$, $C^\mathrm{O}$, and
$\varphi^\mathrm{O}_0$ at given $\mu_{e0}^\mathrm{I}$ and
$\mu_{e0}^\mathrm{O}$. The radius $R$ is fixed based on the dynamic stability of the quark-hadron interface, i.e.,
\begin{equation}
P^\mathrm{I}(R) -2\frac {\sigma}{R} = P^\mathrm{O}(R). \label{eq:P_stable}
\end{equation}
The Wigner-Seitz cell radius $R_\mathrm{W}$ is obtained by minimizing the energy per baryon $M/A$ at a given baryon number density
$n=A/V_\mathrm{W}$ ($V_\mathrm{W}=4\pi R_\mathrm{W}^3/3$), where the total mass $M$ and baryon number $A$ are fixed with
\begin{eqnarray}
M &=& \int_{V_\mathrm{W}} \left[ E(r)+ \frac {1}{8\alpha\pi} \left(\frac{\mbox{d}\varphi}{\mbox{d}r}\right)^2 \right] \mbox{d}V + 4\pi R^2\sigma,
\label{eq:mass_tot}\\
A &=& \int_{V_\mathrm{W}} n(r)\mbox{d}V. \label{eq:A}
\end{eqnarray}
Note that analytical expressions can be obtained for $M$ and $A$ based on Eqs.~(\ref{eq:E_lin}) and (\ref{eq:n}).

\begin{figure}
\includegraphics[width=\linewidth]{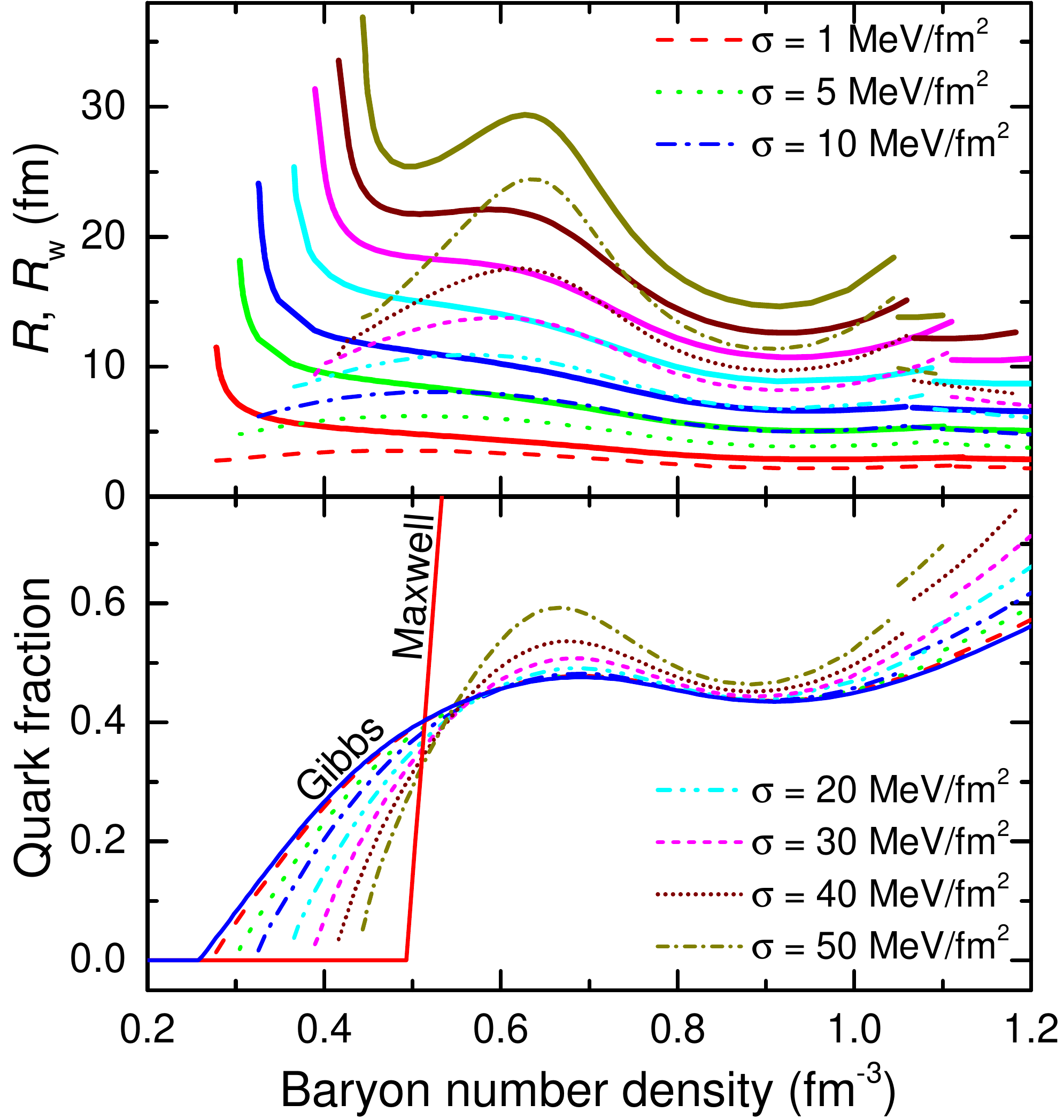}
\caption{\label{Fig:Rfq} Wigner-Seitz cell radius $R_W$ (thick solid curve), droplet radius $R$ (dashed curve), and quark fraction
$\chi$ as functions of baryon number density, where the parameter set $C=2.7$, $\Delta\mu=800$ MeV, and
$B_\mathrm{QCD} = 400\ \mathrm{MeV\ fm}^{-3}$ is adopted.}
\end{figure}

The properties of MP and the corresponding geometrical structures can then be determined based on Eqs.~(\ref{eq:pdis}-\ref{eq:A}).
To show the interface effects on the properties of MP and hybrid stars, as an example, we take $C=2.7$, $\Delta\mu=800$ MeV,
$B_\mathrm{QCD} = 400\ \mathrm{MeV\ fm}^{-3}$, and various surface tension values with $\sigma = 1$, 5, 10, 20, 30, 40, and
$50\ \mathrm{MeV/fm}^{2}$. The obtained Wigner-Seitz cell radius $R_W$, droplet radius $R$, and quark fraction $\chi$ are presented
in Fig.~\ref{Fig:Rfq} as functions of baryon number density $n$. The quark droplet starts to appear in nuclear matter at around
$2n_0$ with a large Wigner-Seitz cell radius $R_W$ and small droplet radius $R$. As density increases, the thickness of the nuclear
matter shell $R_W-R$ decreases, corresponding to the increasing quark fraction $\chi$. {Nevertheless, the quark fraction
$\chi$ starts to decrease at $\sim0.65\ \mathrm{fm}^{-3}$ and reaches the minimum at $\sim0.9\ \mathrm{fm}^{-3}$. This is mainly due
to the fact that we have adopted stiff EoSs for quark matter, which is necessary for hybrid stars to be heavy as $2\ M_\odot$.}
Note that in a large density range the
droplet phase is more stable, while the bubble phase starts to take place at much larger densities ($\sim1\ \mathrm{fm}^{-3}$).
As the surface tension $\sigma$ increases, the sizes of geometrical structures increase and the density range of MP decreases
with an increasing onset density for QM, which is consistent with previous findings. According to the obtained quark fractions,
our results become closer to the cases of the Gibbs construction if we adopt smaller $\sigma$, while at larger $\sigma$ they
approach to the cases of the Maxwell construction.

\begin{figure}
\includegraphics[width=\linewidth]{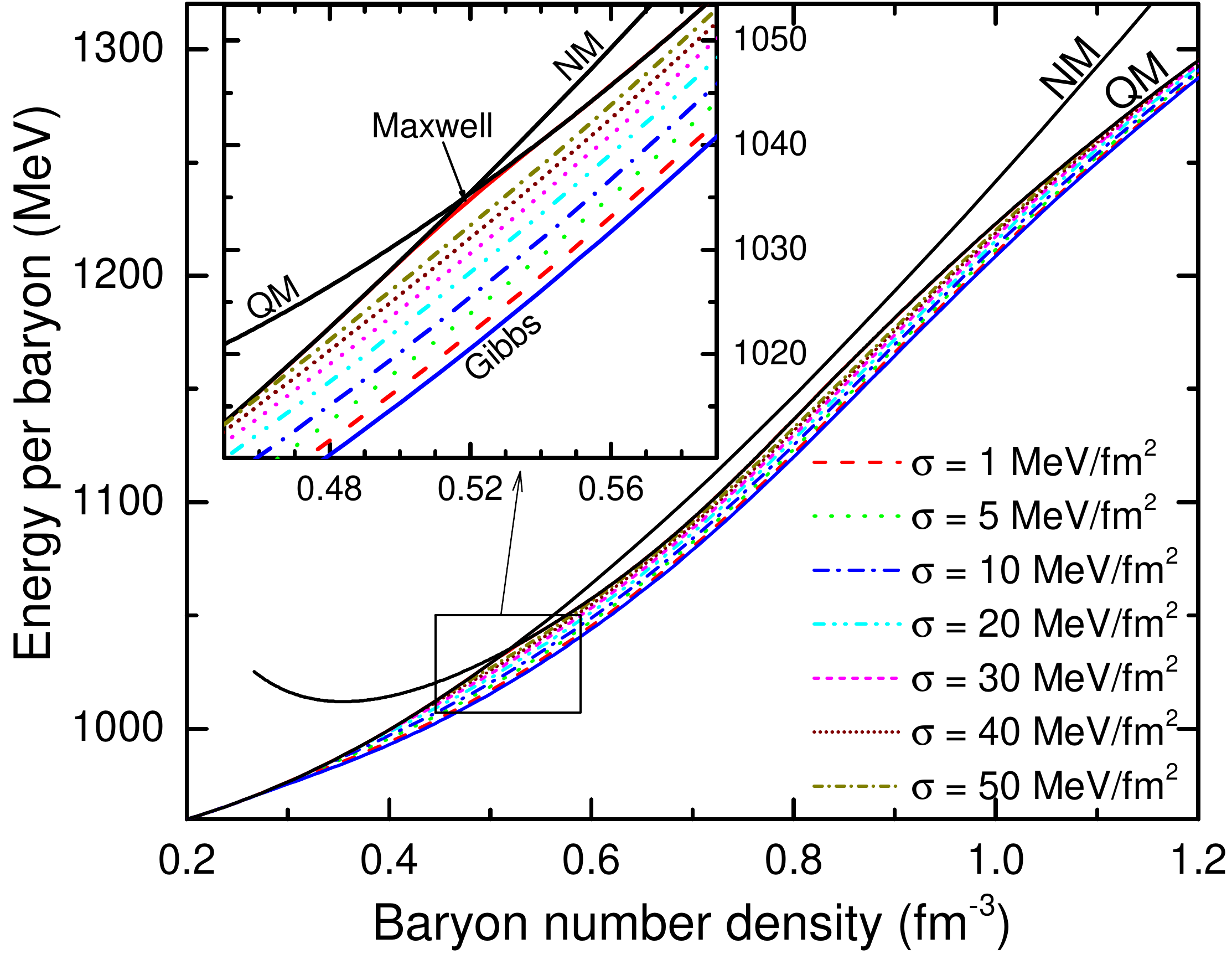}
\caption{\label{Fig:EoS} {Energy per baryon} of hybrid star matter obtained with various quark-hadron interface tensions,
where the parameter set $C=2.7$, $\Delta\mu=800$ MeV, and $B_\mathrm{QCD} = 400\ \mathrm{MeV\ fm}^{-3}$ is adopted.}
\end{figure}

In Fig.~\ref{Fig:EoS} we present the {energy per baryon} of nuclear matter (NM), QM, and MP in compact stars
{as functions of baryon number density}. According to Fig.~\ref{Fig:Rfq},
the onset density for QM increases with $\sigma$, which approaches to the transition densities $n^H_\mathrm{T}$ obtained with
the Maxwell construction. Similarly, the corresponding energy density for the occurrence of deconfinement phase transition
increases with $\sigma$. As the emergence of QM in NM, the EoSs become softer, which will consequently affect the properties
of hybrid stars.

\begin{figure}
\includegraphics[width=\linewidth]{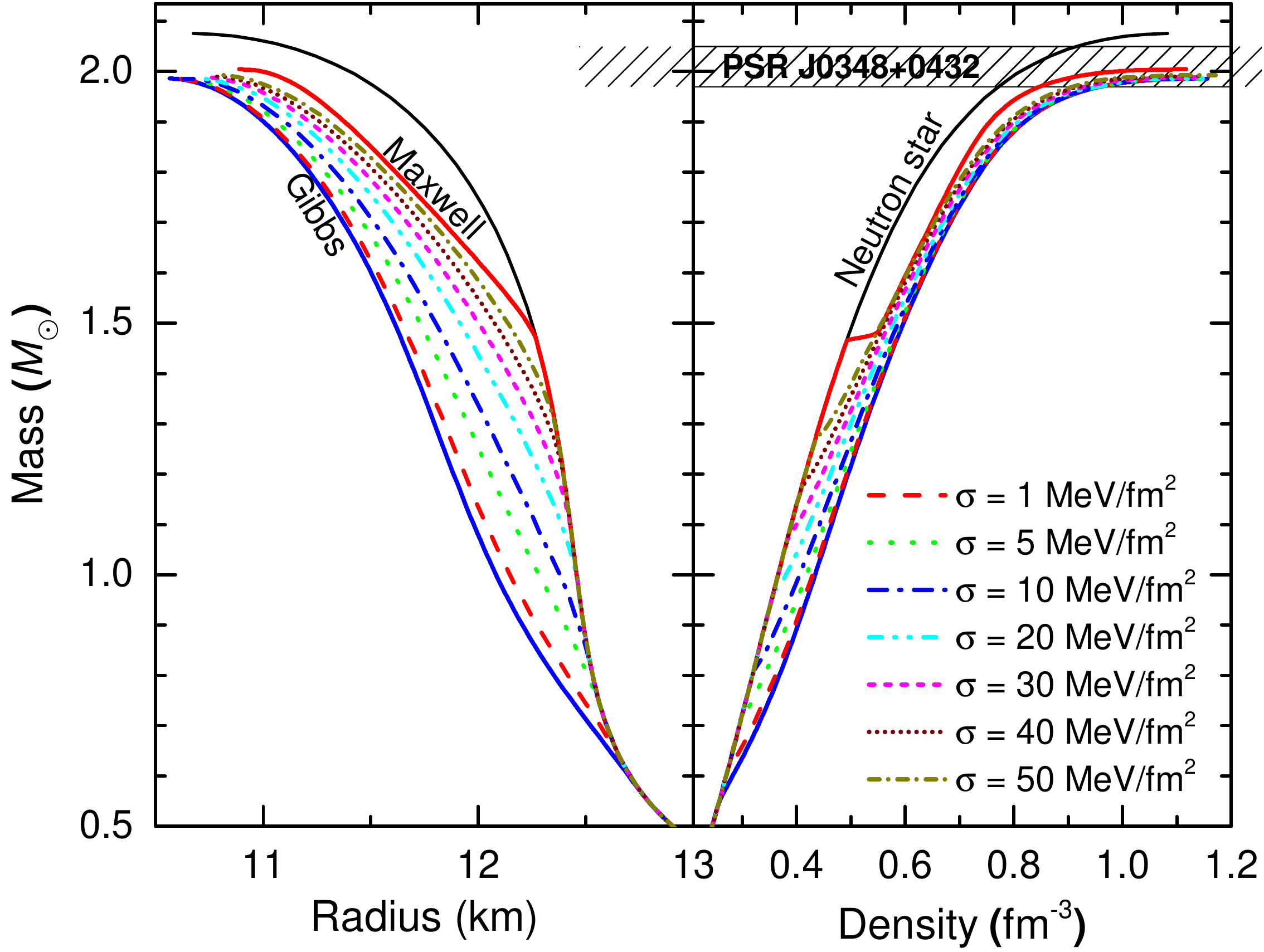}
\caption{\label{Fig:MR} Mass-radius relations of hybrid stars obtained with various quark-hadron interface tensions.
The mass of PSR J0348+0432 ($2.01 \pm 0.04\ M_\odot$)~\cite{Antoniadis2013_Science340-1233232} is indicated with the horizonal band.}
\end{figure}

Based on the EoSs indicated in Fig.~\ref{Fig:EoS}, we solve the TOV equation~(\ref{eq:TOV}) and obtain the structures of
compact stars. In Fig.~\ref{Fig:MR} we present the masses of compact stars as functions of radius (Left panel) and central baryon
number density (Right panel), which are compared with the observational mass of PSR J0348+0432 ($2.01 \pm 0.04\ M_\odot
$)~\cite{Antoniadis2013_Science340-1233232}. As QM starts to appear at the centre of hybrid stars, the maximum mass and radii
become smaller. Comparing with the variations caused by introducing different $C$, $\Delta\mu$, and $B_\mathrm{QCD}$,
interface effects on the maximum mass of hybrid stars are insignificant, which is consistent with our findings in Fig.~\ref{Fig:DMmax_B}
by introducing both the Gibbs and Maxwell constructions. Nevertheless, the interface effects on hybrid stars' radii are sizable,
where the variations can be as large as 600 m. For hybrid stars with a given mass, the radius increases with $\sigma$.

\begin{figure}
\includegraphics[width=\linewidth]{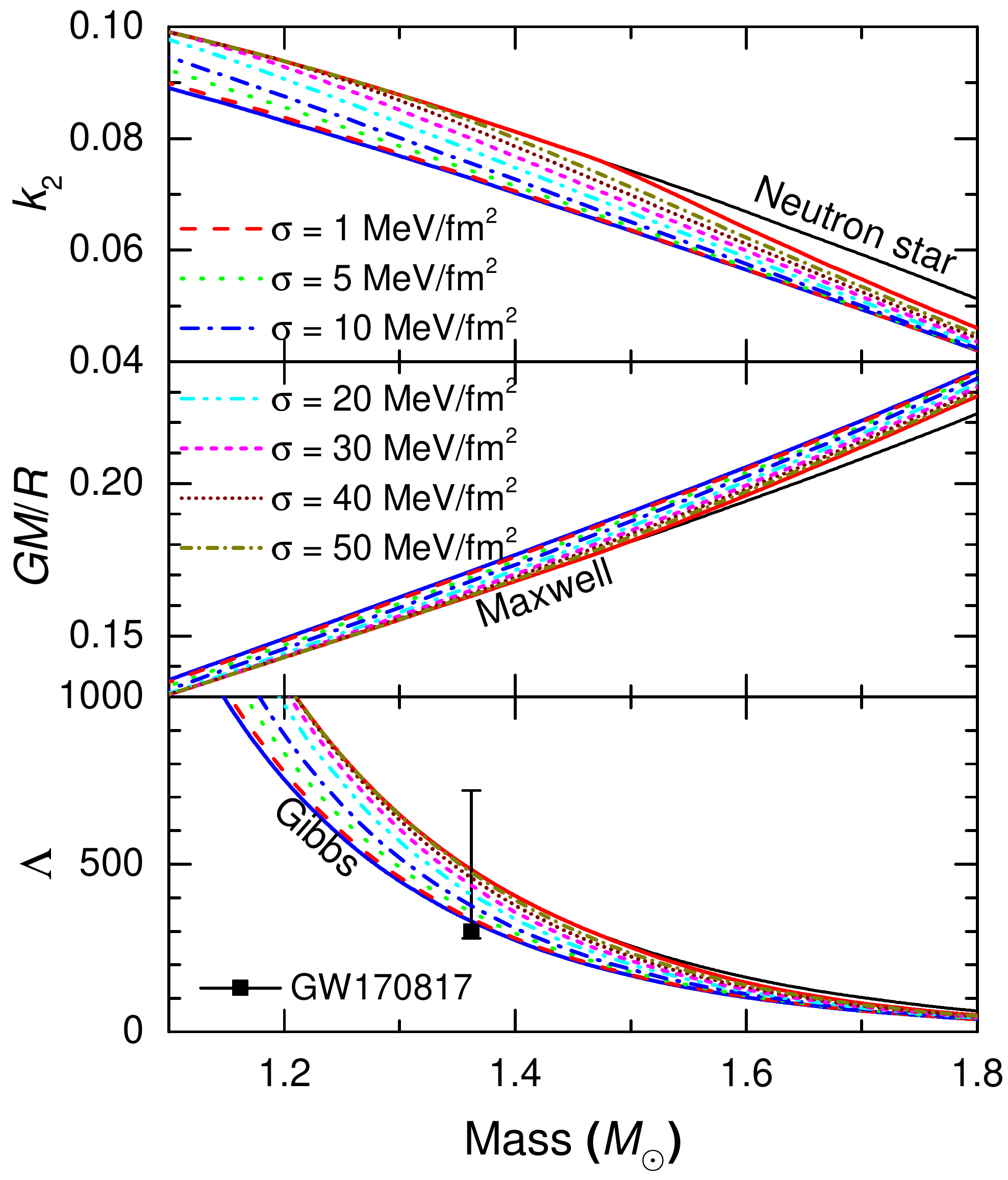}
\caption{\label{Fig:Lambda} {The Love number $k_2$, compactness $GM/R$, and tidal deformability $\Lambda$}
of hybrid stars as functions of their masses. The recent constraint obtained
with the binary neutron star merger event GW170817 is indicated with the black solid box~\cite{LVC2017_PRL119-161101, LVC2019_PRX9-011001,
Coughlin2018_arXiv1812.04803}.}
\end{figure}

\begin{figure}
\includegraphics[width=\linewidth]{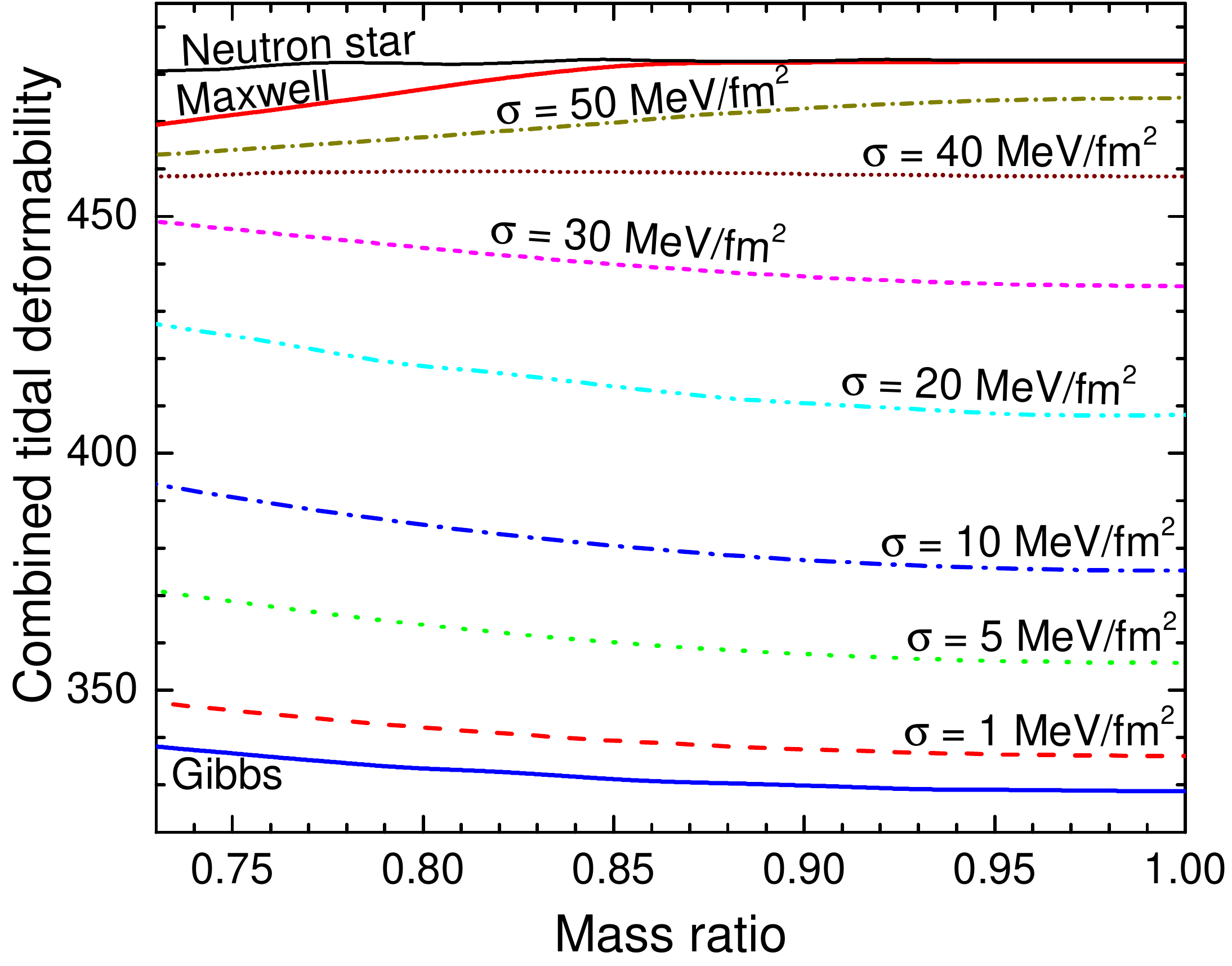}
\caption{\label{Fig:LambdaT} The dimensionless combined tidal deformabilities of hybrid stars as functions of their mass ratio,
which are obtained based on the tidal deformabilities presented in Fig.~\ref{Fig:Lambda} and the chirp mass $\mathcal{M}=1.186\
M_\odot$ of the binary neutron star merger event GW170817~\cite{LVC2019_PRX9-011001}. }
\end{figure}

The tidal deformabilities of compact stars corresponding to Fig.~\ref{Fig:MR} can be obtained with Eq.~(\ref{eq:td}),
which are presented in Fig.~\ref{Fig:Lambda} {along with the corresponding Love number $k_2$ and compactness $GM/R$.
It is found that the tidal deformability increases if we adopt larger $\sigma$, with the positive contributions from both the
increasing $k_2$ and decreasing $GM/R$.} Based on the observations of the binary neutron star merger
event GW170817 and its transient counterpart AT2017gfo and short gamma ray burst GRB170817A, the dimensionless combined tidal
deformability is constrained within $279\leq \tilde{\Lambda} \leq 720$~\cite{LVC2017_PRL119-161101, LVC2019_PRX9-011001,
Coughlin2018_arXiv1812.04803, Carney2018_PRD98-063004, De2018_PRL121-091102, Chatziioannou2018_PRD97-104036}, which
is a mass-weighted linear combination of tidal deformabilities~\cite{Favata2014_PRL112-101101}
\begin{equation}
  \tilde{\Lambda} = \frac{16}{13} \frac{(m_1+12 m_2)m_1^4\Lambda_1+(m_2+12 m_1)m_2^4\Lambda_2}{(m_1+m_2)^5}.
\end{equation}
With the best measured chirp mass $\mathcal{M} = {(m_1 m_2)^{3/5}}{(m_1+m_2)^{-1/5}}=1.186\pm 0.001\
M_\odot$~\cite{LVC2019_PRX9-011001}, in Fig.~\ref{Fig:LambdaT} we present the obtained $\tilde{\Lambda}$ as functions
of the mass ratio $m_2/m_1$. Unlike {traditional neutron stars}~\cite{Bhat2018_JPG46-014003}, a slight deviation
($|\Delta\tilde{\Lambda}|\lesssim 20$) is observed for hybrid stars as we vary $m_2/m_1$, which is insignificant
comparing with the deviations caused by the interface effects. Similar cases are also observed for other choices
of parameters, where the deviation on $\tilde{\Lambda}$ is insignificant. It is then convenient for us to assume
$m_1=m_2=1.362\ M_\odot$ with $\Lambda_1=\Lambda_2=\tilde{\Lambda}$. The corresponding constraints are indicated
in Fig.~\ref{Fig:Lambda}, which can not distinguish hybrid stars obtained at different $\sigma$. Nevertheless,
the situation will likely be changed in the near future with the implementation of {upgraded detectors}
for gravitational wave observations~\cite{Torres-Rivas2019_PRD99-044014, Bauswein2019_arXiv1901.06969}.

\begin{figure}
\includegraphics[width=\linewidth]{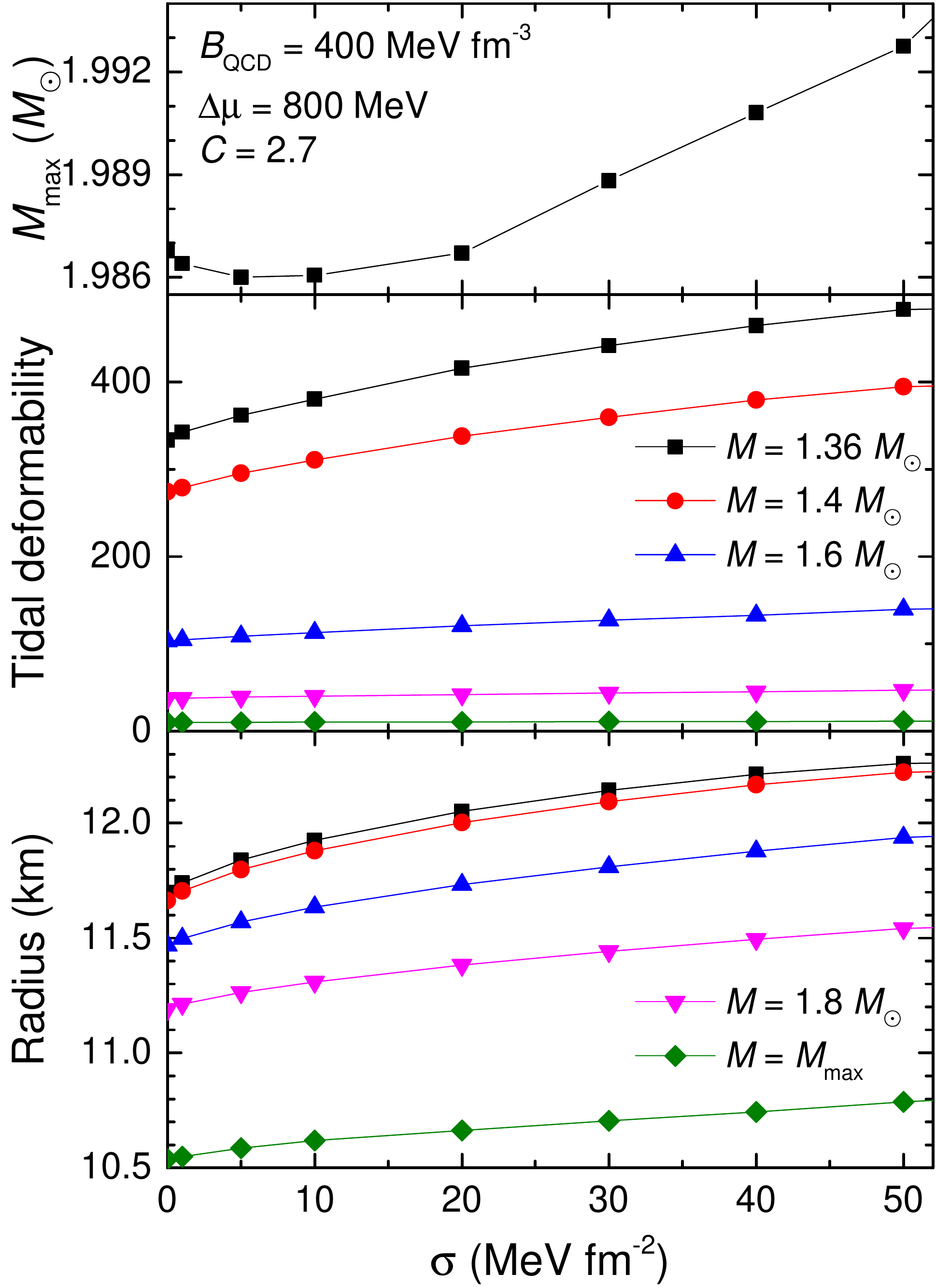}
\caption{\label{Fig:MLRSigma} The evolution of maximum mass (top panel), tidal deformability (centre panel), and radius (bottom panel)
of hybrid stars as functions interface tension.}
\end{figure}

Based on Figs.~\ref{Fig:MR} and~\ref{Fig:Lambda}, in Fig.~\ref{Fig:MLRSigma} we present the evolution of
maximum mass, tidal deformability, and radius of hybrid stars as functions of the surface tension $\sigma$, with
given masses $M=1.36$, 1.4, 1.6, $1.8\ M_\odot$ and $M_\mathrm{max}$. For the cases with $\sigma=0$, we adopt
the results obtained with the Gibbs construction. It is found that $M_\mathrm{max}$, $\Lambda$, and $R$ increase
with $\sigma$. The results essentially interpolate between two types of values, i.e., as a function of $\sigma$
that connects the results obtained with the Gibbs construction at $\sigma\rightarrow 0$ and the Maxwell construction
at $\sigma>\sigma_\mathrm{c}$, where we have found $\sigma_\mathrm{c} = 79.12$ MeV/fm${}^2$. In principle, the
corresponding function can be obtained by fitting to our results with certain assumption on its form, e.g.,
\cite[Eq.~(14)]{Maslov2018_arXiv1812.11889}, which we intend to do in our future works. According to
Fig.~\ref{Fig:MLRSigma}, we find the variations on the maximum mass, tidal deformability, and radius of hybrid
stars are up to $\Delta M_\mathrm{max} \approx 0.02\ M_\odot$, $\Delta R \approx 600$ m, and $\Delta \Lambda/\Lambda
\approx 50\%$, respectively. Even though $M_\mathrm{max}$ increases little with $\sigma$, sizable changes are
observed for $\Lambda$ and $R$, which is within the capability of the NICER mission~\cite{Gendreau2016_PSPIE9905-16}
or future gravitational wave observations~\cite{Torres-Rivas2019_PRD99-044014, Bauswein2019_arXiv1901.06969}.

\begin{figure}
\includegraphics[width=\linewidth]{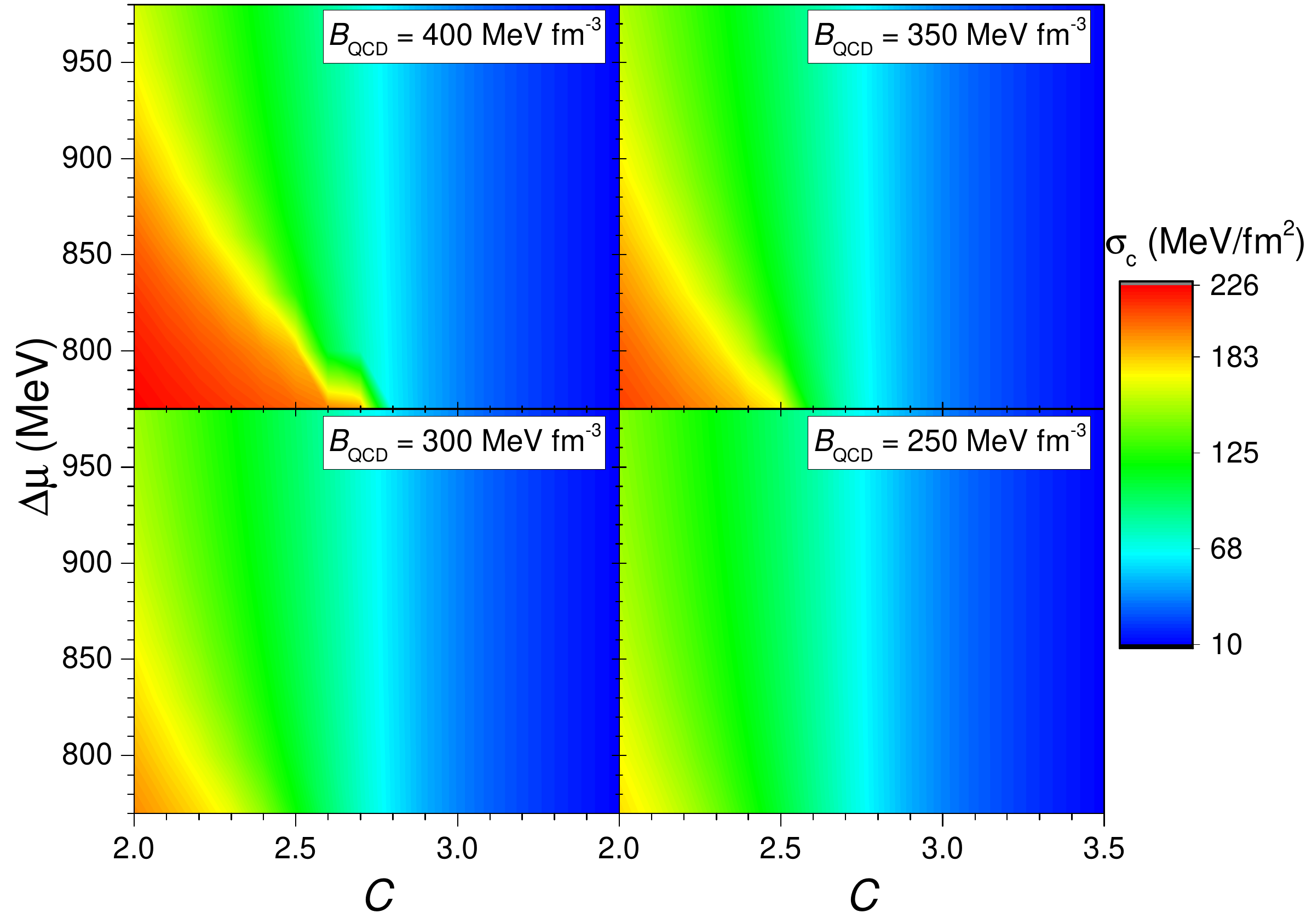}
\caption{\label{Fig:SigmaC_B} The critical surface tension $\sigma_\mathrm{c}$ estimated with Eq.~(\ref{eq:sigma_c}).}
\end{figure}

\begin{figure}
\includegraphics[width=\linewidth]{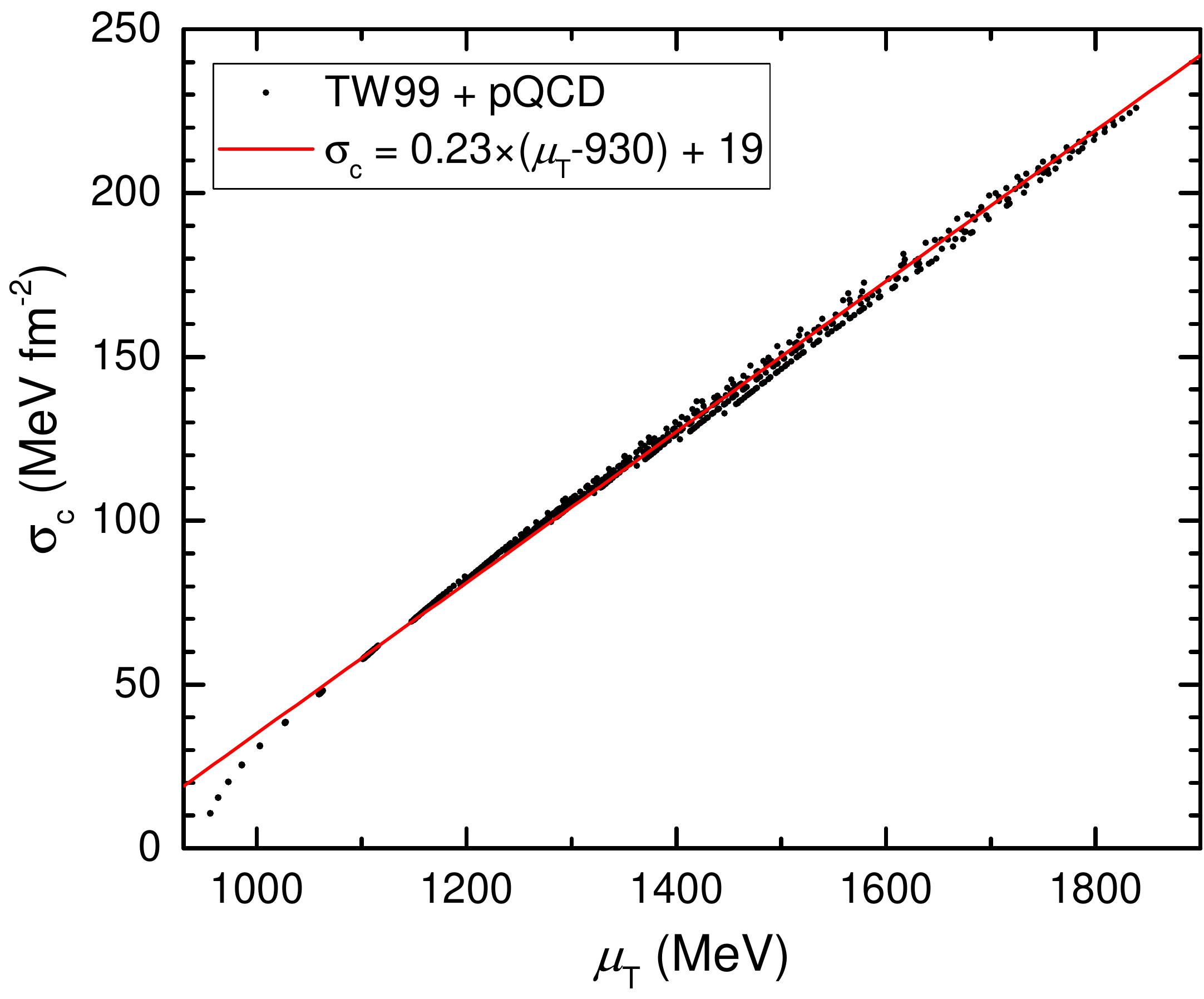}
\caption{\label{Fig:SigmaCmuT} The critical surface tension $\sigma_\mathrm{c}$ as a function of the chemical potential $\mu_T$
on the occurrence of deconfinement phase transition.}
\end{figure}

Finally, our investigations in Fig.~\ref{Fig:EoS}-\ref{Fig:MLRSigma} suggests that the radii and tidal deformabilities
of hybrid stars are monotonous functions of $\sigma$, which approach to the scenarios of Gibbs construction at
$\sigma\rightarrow 0$ and Maxwell construction at $\sigma>\sigma_\mathrm{c}$. For other choices of parameter sets,
we expect similar trends, where the critical surface tension $\sigma_\mathrm{c}$ can be well reproduced
with~\cite{Voskresensky2003_NPA723-291}
\begin{equation}
\sigma_\mathrm{c} = \frac{\left( \mu_{e0}^H - \mu_{e0}^Q \right)^2}{8 \pi \alpha \left(\lambda_D^Q + \lambda_D^H\right)}.
\label{eq:sigma_c}
\end{equation}
The obtained results for $\sigma_\mathrm{c}$ are presented in Fig.~\ref{Fig:SigmaC_B}, where we have observed similar trends
as in Fig.~\ref{Fig:nHT}.
This indicates correlations between the critical surface tension $\sigma_\mathrm{c}$ and the thermodynamic quantities of MP,
which was pointed out in \cite[Fig.~5]{Maslov2018_arXiv1812.11889}. To show this explicitly, in Fig.~\ref{Fig:SigmaCmuT}
we present the obtained critical surface tension as a function of the baryon chemical potential $\mu_T$ on the occurrence
of deconfinement phase transition. It is found that $\sigma_\mathrm{c}$ (in MeV/fm${}^2$) increases linearly with $\mu_T$
(in MeV) and can be well approximated with $\sigma_\mathrm{c} = 0.23(\mu_T-930)+19$, where the coefficients depend on the
EoSs of HM and QM. In such cases, if the deconfinement phase transition occurs at large $\mu_T$, the emergence of geometrical
structures may be inevitable in hybrid stars since typical estimations suggest $\sigma<\sigma_\mathrm{c}$ with $\sigma
\lesssim 30\ \mathrm{MeV/fm}^{2}$~\cite{Oertel2008_PRD77-074015, Palhares2010_PRD82-125018, Pinto2012_PRC86-025203,
Kroff2015_PRD91-025017, Garcia2013_PRC88-025207, Ke2014_PRD89-074041, Mintz2013_PRD87-036004, Gao2016_PRD94-094030,
Xia2018_PRD98-034031}.

\section{\label{sec:con}Conclusion}
We investigate the interface effects of quark-hadron mixed phase in hybrid stars. The properties of nuclear matter are obtained
based on RMF model.  For the $N$-$N$ interactions, we adopt the covariant density functional TW99~\cite{Typel1999_NPA656-331},
which is consistent with all seven constraints related to symmetric nuclear matter, pure neutron matter, symmetry energy, and
its derivatives~\cite{Dutra2014_PRC90-055203}. For the quark phase, we adopt perturbation model by expanding the pQCD thermodynamic
potential density to the order of $\alpha_\mathrm{s}$~\cite{Fraga2005_PRD71-105014}. A parameterized bag constant is
introduced by comparing with pQCD calculations to the order of $\alpha_\mathrm{s}^2$~\cite{Fraga2014_ApJ781-L25} as well as
incorporating informations from QCD sum-rule~\cite{Shuryak1978_PLB79-135} and light hadron mass spectra~\cite{DeGrand1975_PRD12-2060}.
Since the mixed phases obtained with the Gibbs and Maxwell constructions correspond to the two limits of the quark-hadron interface
tension, i.e.,  $\sigma\rightarrow 0$ for the Gibbs construction and $\sigma>\sigma_\mathrm{c}$ for the Maxwell construction,
we investigate the interface effects by comparing the results obtained by those two phase construction schemes. It is found that
the quark-hadron interface has sizable effects on the radii ($\Delta R \approx 600$ m) and tidal deformabilities ($\Delta
\Lambda/\Lambda \approx 50\%$) of {1.36 solar mass} hybrid stars for certain choices of parameters. This is then confirmed by considering the
geometrical structures of the mixed phase with a specific choice of parameters, where for larger $\sigma$ the sizes of geometrical
structures increase but the density range of mixed phase decreases. For the corresponding hybrid stars, we find the maximum mass,
tidal deformability, and radius increase with $\sigma$, where the variations are up to $\Delta M_\mathrm{max} \approx 0.02\ M_\odot$,
$\Delta R \approx 600$ m, and $\Delta \Lambda/\Lambda \approx 50\%$, respectively. This provides possibilities for us to
constrain the quark-hadron interface tension with future gravitational wave observations~\cite{Torres-Rivas2019_PRD99-044014,
Bauswein2019_arXiv1901.06969} as well as the NICER mission~\cite{Gendreau2016_PSPIE9905-16}.

\section*{ACKNOWLEDGMENTS}
NY, TM, and TT would like to thank K. Maslov, A. Ayriyan, D. Blaschke, H. Grigorian, and D. N. Voskresensky for fruitful
discussion. This work was supported by National Natural Science Foundation of China (Grant Nos.~11705163, 11711540016,
11875052, 11575190, and 11525524), and JSPS KAKENHI (Grant No.~17K18792).
The computation for this work was supported by the HPC Cluster of SKLTP/ITP-CAS and the Supercomputing Center, CNIC, of the CAS.

%%
%% reference here
%%
%\bibliography{strange_quark}


\end{document}